\documentstyle[12pt,epsf]{article}

\newcommand{\journal}[4]{{\rm #1} {\bf #2} (19#3) #4}
\newcommand{\NP}{\journal{Nucl. Phys.}}
\newcommand{\PL}{\journal{Phys. Lett.}}
\newcommand{\PRL}{\journal{Phys. Rev. Lett.}}
\newcommand{\NPPS}{\journal{Nucl. Phys. Proc. Suppl.}}
\newcommand{\MPCPS}{\journal{Math. Proc. Camb. Phil. Soc.}}

\makeatletter
  
  \@addtoreset{equation}{section}
\makeatother

\begin{document}

\begin{titlepage}

\begin{flushright}
OU-HET 295 \\
April 1998
\end{flushright}
\bigskip
\bigskip

\begin{center}
{\Large \bf
On $N=2$ $M$QCD
}${}^\dagger$

\bigskip
Toshio Nakatsu\\
\bigskip
{\small \it
Department of Physics,\\
Graduate School of Science, Osaka University,\\
Toyonaka, Osaka 560, JAPAN
}
\end{center}
\bigskip
\bigskip
\begin{abstract}
We review $M$-theory description 
of $4d$ $N\!=\!2$ SQCD. 
Configurations of $M$-theory fivebranes  
relevant to describe  
the moduli spaces of the Coulomb and Higgs 
branches are studied using the Taub-NUT geometry. 
Minimal area membranes related with the BPS 
states of $N\!=\!2$ SQCD are given explicitly. 
They almost saturate the BPS bounds. 
The deviation from the bounds is due to 
their boundary condition constrained by 
the fivebrane. 
The electric-magnetic duality 
at the baryonic branch root 
is also examined from the $M$-theory viewpoint. 
In this course, 
novel concepts such as creation of brane 
and exchange of branes in Type II theory 
are explained in the framework 
of $M$-theory.

\end{abstract}

\vfill 
\hrule 
\vskip 3mm 
\begin{small} 
\noindent{$\dagger$}
Lecture in the Second Winter School on 
``Branes, Fields And Mathematical Physics" 
at the APCTP (Feb.9-20, 1998).
\end{small}

\end{titlepage}


    Many interesting results about gauge field theories 
in various dimensions have been obtained by analyzing 
the worldvolume effective theory of branes in superstring 
theory \cite{GK}. 
They can be realized by branes mostly in Type IIA or IIB 
superstring theory, 
but paticularly an interesting configuration, 
which describes four-dimensional $N\!=\!2$ supersymmetric QCD 
($4d$ $N\!=\!2$ SQCD), has been proposed 
in \cite{Witten1} whithin the framework of $M$-theory. 
In this construction a mysterious hyper-elliptic 
curve, which is used for the description of the exact 
solution of the Coulomb branch of $4d$ $N\!=\!2$ SQCD, 
appears as a part of a $M$-theory fivebrane. 

      It is pointed out 
\cite{SW1,SW2,S,IS} that there exist  
various dualities between supersymmetric gauge 
field theories and 
that these dualities play an important 
role for our understanding of their non-perturbative 
dynamics.  
Several steps to clarify an origin of these dualities 
from the string theory viewpoint have been taken 
\cite{HW,EGK}. In this course of explanation 
we need novel concepts such as a brane can be created 
or annihilated  
when two different branes cross each other. 
However, since these phenomena are due to the strong 
coupling dynamics of Type II theories, it is still 
difficult to treat them correctly in these theories. 
On the other hand, $M$-theory includes the strong coupling 
dynamics of Type IIA theory in its semi-classical 
description \cite{Witten2}. 
So one can expect that the ``dualities" accompanied by 
exchanging branes can be understood via semi-classical 
analysis of $M$-theory.

           In this article we review 
$M$-theory description of $4d$ $N\!=\!2$ SQCD. 
Configurations of $M$-theory fivebranes  
relevant to describe the moduli spaces 
of the Coulomb and Higgs branches 
are studied using the Taub-NUT geometry. 
Minimal area membranes related with the BPS 
states of $N\!=\!2$ SQCD are given explicitly, 
but, due to their boundary condition, 
they can not saturate the BPS bounds 
at a generic point of the moduli space 
of the configuration.
This signals a necessity 
to incorporate some recoil of the 
membranes to the fivebrane. 
The electric-magnetic duality 
at the baryonic branch root of 
$4d$ $N\!=\!2$ SQCD 
is also examined from the $M$-theory viewpoint. 
In this course, 
novel concepts such as creation of brane 
and exchange of branes in Type II theory 
are explained in the framework 
of $M$-theory.

        The organization is as follows : 
In section 1 we start with 
a realization of the Coulomb 
branch of $4d$ $N\!=\!2$ SQCD  
by $M$-theory fivebrane configuration. 
In this description 
the worldvolume of fivebrane includes 
the so-called Seiberg-Witten curve, as a part. 
An inclusion of matters is expressed 
by an embedding of the curve into 
the multi-Taub-NUT space. 
This embedding is studied in detail 
by using the concrete metric of 
the multi-Taub-NUT space. 
In section 2 minimal area membranes 
are studied treating the previous $M$-theory 
brane configuration 
as their background geometry. 
These membranes are classified 
by their shapes in ten-dimensions and 
boundary conditions in eleven-dimensions. 
In section 3 we study the Higgs branch 
of $4d$ $N\!=\!2$ SQCD. 
An exact description of 
the baryonic and non-baryonic branches 
in terms of $M$-theory is presented. 
The $s$-rule \cite{HW} is also derived 
using the Taub-NUT geometry. 
In the last section 
the baryonic branch root, 
that is, 
iintersection of the Coulomb branch 
and the baryonic branch 
is examined from the $M$-theory viewpoint. 
It is known \cite{APS} in field theory that 
the baryonic branch root admits to have 
two descriptions dual to each other. 
After providing $M$-theory proof 
of  ``brane creation",  
we present a specific family of $M$-theory 
configurations relevant 
to explain this duality and examine 
the BPS states of membranes 
associated with them.


\section{Coulomb Branch of $N\!=\!2$ MQCD}

         Four-dimensional gauge theories 
with $N\!=\!2$ supersymmetry can be realized 
\cite{Witten1} as effective theories 
on the world-volume of a  $M$-theory fivebrane.  
A different type of gauge theories requires 
a different topology of $M$-theory fivebrane 
and a different eleven-dimensional background 
where the fivebrane is embedded. 
This realization of $N\!=\!2$ supersymmetric QCD 
via the world-volume effective theory 
of the $M$-theory fivebrane is called 
$N\!=\!2$ $M$-theory QCD. (MQCD for short.) 
MQCD does not exactly coincide with 
an ordinary supersymmetric QCD in four-dimension, 
but is considered to belong to the same universality 
class. Moreover many difficulties appearing 
in the field theoretical analysis of 
the supersymmetric QCD vacua, which are 
mainly due to their singularities, 
are resolved within the framework of $M$-theory. 
So, MQCD is a very useful tool for 
our understanding of the dynamics of 
supersymmetric QCD.

                  Consider an eleven-dimensional 
manifold $M^{1,10}$ of $M$-theory 
which admits to have the form 
\begin{eqnarray}
M^{1,10} \simeq {\bf R}^{1,3} \times X^{7}.
\label{decomposition}
\end{eqnarray}
${\bf R}^{1,3}$ is 
the four-dimensional space-time 
where $N\!=\!2$ supersymmetric theory will exist. 
$X^7$ is a (non-compact) 
seven-dimensional manifold 
which suffers several constraints due to 
the requirement of $N\!=\!2$ supersymmetry 
on the worldvolume. 
The supersymmetry of $M$-theory 
in the eleven-dimensions is 
generally broken by 
this product space structure of $M^{1,10}$.
However, 
if the submanifold $X^7$ admits to have 
a holonomy group smaller than $SO(7)$, 
some of supersymmetries will survive 
on the four-dimensional space-time ${\bf R}^{1,3}$. 
Recall that we ultimately 
realize $N\!=\!2$ supersymmetry 
on the worldvolume of a $M$-theory fivebrane, 
strictly speaking, 
on its four-dimensional part which is 
identified with ${\bf R}^{1,3}$ 
in (\ref{decomposition}). 
The fivebrane itself will be introduced 
soon later as a BPS saturated state 
which breaks half 
the surviving supersymmetries. 
So, with this reason, we must take $X^7$ as 
a submanifold which keeps $N=4$ supersymmetry 
on the four-dimensional space-time 
${\bf R}^{1,3}$. 
The holonomy group of $X^7$ is required 
to be $SU(2)$. 
This requirement reduces 
the seven-manifold $X^7$ to be
\begin{eqnarray}
X^7 \simeq {\bf R}^3 \times Q^4,
\label{X7}
\end{eqnarray}
where $Q^4$ is a four-manifold 
with $SU(2)$ holonomy, that is, 
a hyper-K\"ahler manifold.  
This hyper-K\"ahler manifold should be chosen 
appropriately according to whether the theory 
on ${\bf R}^{1,3}$ 
contains matter hypermultiplets or not.  


\subsection{Pure $N\!=\!2$ MQCD}

               Let us describe a configuration of 
$M$-theory fivebrane suitable to our purpose. 
Since we want to leave $N\!=\!2$ supersymmetry 
on ${\bf R}^{1,3}$ as the supersymmetry 
of the worldvolume effective theory of fivebrane, 
the worldvolume itself must fill 
all of ${\bf R}^{1,3}$. 
The rest of the fivebrane is 
a two-dimensional surface 
$\Sigma$ in $X^7$. 
The Lorentz group $SO(3)$ of ${\bf R}^3$ 
in (\ref{X7}) turns out to be 
the $SU(2)_R$-symmetry of 
$N\!=\!2$ supersymmetry algebra in four-dimensions. 
In order to preserve this symmetry 
$\Sigma$ must lie at a single point in ${\bf R}^3$.
It can only spread in $Q^4$ 
as a two-dimensional surface. 
To summarize, 
the worldvolume of the fivebrane 
must be ${\bf R}^{1,3}\times\Sigma$ 
where ${\bf R}^{1,3}$ is identified with 
the four-dimensional space-time 
and $\Sigma$ is a two-dimensional surface 
embedded in $Q^4$.  

              Further restriction 
on the fivebrane world-volume is that 
it must be BPS-saturated in order to preserve 
a half of supersymmetries. 
This is achieved by the minimal area 
embedding of $\Sigma$ into $Q^4$. 
The area $A_\Sigma$ of the surface $\Sigma$ 
is bounded from the below 
\begin{eqnarray}
A_\Sigma \geq \int_\Sigma \Omega_{\bf R},
\end{eqnarray}
where $\Omega_{\bf R}$ is the K\"ahler form 
of the hyper-K\"ahler manifold $Q^4$. 
This inequality is saturated 
if and only if $\Sigma$ 
is holomorphically embedded in $Q^4$. 
This means that, 
if one introduces the holomorphic 
coordinates $y$ and $v$ of $Q^4$, 
the surface $\Sigma$ is a curve in $Q^4$ 
defined by a holomorphic function    
\begin{eqnarray}
F(y,v)=0.
\end{eqnarray}

           Let $\Sigma$ be a Riemann surface 
with genus $N_c\!-\!1$. 
Then there appear $N_c\!-\!1$ massless 
vector multiplets 
on ${\bf R}^{1,3}$\cite{Verlinde}. 
So, the effective theory on ${\bf R}^{1,3}$ is 
a  supersymmetric $U(1)^{N_c-1}$ gauge theory. 
In particular let $\Sigma$ be  
the Seiberg-Witten curve $\Sigma_{SW}$ \cite{SW1,SW2} 
for the pure $SU(N_c)$ gauge theory \cite{KLY,AF}. 
\begin{eqnarray}
y^2-2\prod_{a=1}^{N_c}(v-\phi_a)y+\Lambda^{2N_c}=0.
\label{pure}
\end{eqnarray}

       We now take, 
as the four-manifold $Q^4$, a flat space
$Q_0 \! \equiv \! {\bf R}^3\times S^1$ 
with coordinates 
$(x^4,x^5,x^6,x^{10})$. 
$S^1$ is the circle of the eleven-dimension 
with radius $R$. 
For the later convenience 
we rescale the coordinates 
and define 
\begin{eqnarray}
(v,b,\sigma) \equiv 
\frac{2}{R}(x^4+ix^5,x^6,x^{10}). 
\label{v b sigma}
\end{eqnarray} 
The complex structure appropriate 
to describe the holomorphic embedding 
of the Seiberg-Witten curve   
is given \cite{Witten1} 
by $v$ and 
$y(b,\sigma)\!=\!e^{-\frac{b+i \sigma}{2}}$.   
Using these holomorphic coordinates 
the (trivial) hyper-K\"ahler structure of $Q_0$ 
can be written as 
\begin{eqnarray} 
\Omega_{\bf R}&=& 
\frac{i R^2}{2}
\left( \frac{1}{4}dv \wedge d \overline{v}
+ \frac{dy}{y} \wedge \overline{\frac{dy}{y}} 
\right), 
\nonumber \\ 
\Omega_{\bf C}&=& 
\frac{iR^2}{2}dv \wedge \frac{dy}{y},  
\end{eqnarray} 
where $\Omega_{{\bf R},{\bf C}}$ are 
respectively 
K\"ahler and holomorphic two-forms 
\footnote{We normalize $\Omega_{{\bf R},{\bf C}}$ 
so that 
$dvol_{Q_0}
=\frac{1}{2}\Omega_{\bf R}
\wedge \Omega_{\bf R}
=\frac{1}{4}\Omega_{\bf C}
\wedge \overline{\Omega}_{\bf C}$,  
where $dvol_{Q_0}$ is 
the volume form of $Q_0$, 
$dx^4 \wedge dx^5 \wedge dx^6 \wedge dx^{10}$.}.

        Let us comment on 
the holomorphic embedding of 
$\Sigma_{SW}$ into $Q_0$. 
One can visualize it 
by considering eq.(\ref{pure}) 
in terms of $v,b$ and $\sigma$, that is, 
considering it in the real coordinates of $Q_0$. 
Recall the curve is hyper-elliptic. 
Let $\Sigma_{\pm}$ be its two Riemann sheets  
\begin{eqnarray} 
\Sigma_{SW}= \Sigma_{+} \cup \Sigma_{-}. 
\end{eqnarray} 
These Riemann sheets $\Sigma_{\pm}$ are 
embedded into $Q_0$ by the equations 
\begin{eqnarray} 
y_{\pm}(v)=y(b,\sigma) ,
\end{eqnarray} 
where $y_{\pm}(v)$ are two roots of eq.(\ref{pure})
\begin{eqnarray}
y_{\pm}(v)= 
\prod_{a=1}^{N_c}(v-\phi_a)
\pm \sqrt{\prod_{a=1}^{N_c}(v-\phi_a)^2-\Lambda^{2N_c}}, 
\end{eqnarray}
and the holomorphic coordinate $y$ is regarded as 
the function of $b$ and $\sigma$.

             These Riemann sheets are 
patched together in $Q_0$ 
by $N_c$ one-dimensional circles.  
Let us denote them 
$C_{a}$ $(1 \!\leq \! a \! \leq \! N_c)$. 
It holds 
\begin{eqnarray} 
\Sigma_+ \cap \Sigma_- = 
C_1 \cup \cdots \cup C_{N_c} 
~~~~~(\subset Q_0).
\end{eqnarray} 
Each $C_a$ can be regarded 
as the $S^1$ of $Q_0$, 
that is, the circle of eleven-dimension. 
While this, if considering it 
as a one-dimensional cycle of the curve, 
$C_a$ is a cut of the Riemann sheets, 
that is, the $\alpha$-cycle of the curve.

\begin{figure}[t]
\epsfysize=5cm \centerline{\epsfbox{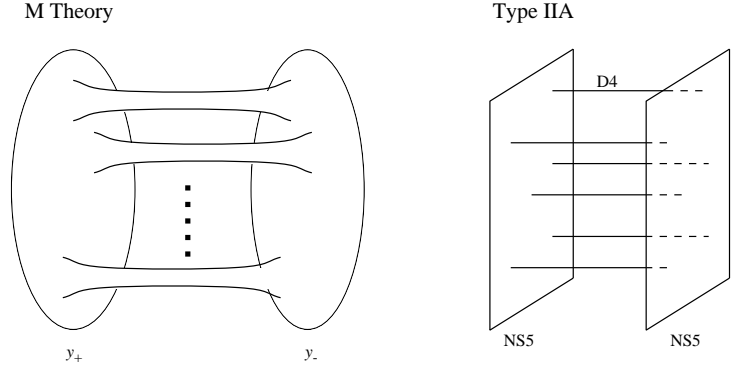}} 
\caption{\small $M$-theory fivebrane and Type IIA configuration.}
\label{MandIIA} 
\end{figure}

           These $N_c$ circles on $\Sigma_{SW}$  
or presumably 
their infinitesimally neighborhoods 
in $\Sigma_{SW}$ 
become $N_c$ D fourbranes in Type IIA theory.  
The residual part of the curve becomes 
two NS fivebranes. 
After all, the single fivebrane in $M$-theory
described by eq.(\ref{pure}) becomes two NS fivebranes with 
worldvolume $(x^0,x^1,x^2,x^3,x^4,x^5)$ and $N_c$ D fourbranes 
with worldvolume $(x^0,x^1,x^2,x^3,[x^6])$\footnote{$[x^6]$ 
describes a finite interval.} stretching between them in the 
$x^6$-direction. 
See Fig.\ref{MandIIA}.

\subsection{$N\!=\!2$ MQCD with matter}

            In Type IIA picture an inclusion of 
$N_f$ matter hypermultiplets into 
the above pure gauge theory 
can be achieved by considering 
$N_f$ D sixbranes with worldvolume 
$(x^0,x^1,x^2,x^3,x^7,x^8,x^9)$ and 
then putting these sixbranes 
between two NS fivebranes.  
In such a configuration there appears 
$N\!=\!2$ $SU(N_c)$ supersymmetric QCD with $N_f$ flavors 
on their common worldvolume $(x^0,x^1,x^2,x^3)$. 
This is because the open string sector 
between $N_c$ D fourbranes and $N_f$ D sixbranes of 
the configuration includes hypermultiplets 
which belong to the fundamental representations 
both of the gauge and flavor groups.

\subsubsection{Sixbranes and the multi-Taub-NUT space}

                    Sixbranes can be regarded 
as the ``Kaluza-Klein monopoles'' of $M$-theory compactified 
to Type IIA theory with a circle $S^1$.  
This is because they are magnetically charged with respect 
to the $U(1)$ gauge field associated 
with this $S^1$.

              Consider a single sixbrane solution 
of Type IIA SUGRA.  
Let us denote 
the worldvolume and transverse directions of 
a sixbrane respectively 
by $\vec{x}_{\|}$ and $\vec{x_{\bot}}$
\footnote{In this article 
$\vec{x}_{\|}
=(x^0,x^1,x^2,x^3,x^7,x^8,x^9)$
and 
$\vec{x}_{\bot}
=(x^4,x^5,x^6)$.}. 
The solution 
\cite{HS} 
can be written by using 
a single potential $V$ 
\begin{eqnarray} 
V(x_{\bot})=1+\frac{g_sl_s/2}{x_{\bot}},  
\end{eqnarray} 
where 
$x_{\bot} \! \equiv \! 
\sqrt{\vec{x}_{\bot}\cdot \vec{x}_{\bot}}$.  
The ten-dimensional metric 
and the dilaton have the forms  
\begin{eqnarray} 
ds_{10}^2 &=&
V(x_{\bot})^{-\frac{1}{2}}
d \vec{x}_{\|} \cdot d \vec{x}_{\|}
+V(x_{\bot})^{\frac{1}{2}}
d \vec{x}_{\bot} \cdot d \vec{x}_{\bot}, 
\nonumber \\ 
e^{\phi}&=& 
V(x_{\bot})^{-\frac{3}{4}}. 
\label{D6 solution in IIA SUGRA} 
\end{eqnarray} 
The $U(1)$ gauge field $A$, 
which has components 
only in the transverse directions, 
is determined by 
\footnote{
$\vec{\nabla}_{x_{\bot}}
\equiv 
(\partial_{x^4},\partial_{x^5},\partial_{x^6})$.}  
\begin{eqnarray}
\vec{\nabla}_{x_{\bot}} \times \vec{A}
= \vec{\nabla}_{x_{\bot}} V(x_{\bot}).  
\end{eqnarray}
Inserting them to  
$ds_{11}^2=$ 
$e^{-\frac{2}{3}\phi}ds_{10}^2$ 
$+e^{\frac{4}{3}\phi}$ 
$\left( dx^{10}+ 
\vec{A}\cdot d \vec{x}_{\bot}\right)^2$, 
we obtain the eleven-dimensional solution :
\begin{eqnarray} 
ds_{11}^2= 
d \vec{x}_{\|}\cdot d \vec{x}_{\|}
+ \left \{
V(x_{\bot}) 
d \vec{x}_{\bot}\cdot d \vec{x}_{\bot}
+\frac{1}{V(x_{\bot})}
\left(dx^{10}
+\vec{A}\cdot d\vec{x}_{\bot} 
\right)^2 
\right \} , 
\label{D6 solution in M}
\end{eqnarray}
where the circle of the eleven-dimension 
has radius $R \!= \! g_sl_s$.
($x^{10} \! \sim \! x^{10} \!+ \! 2\pi R.)$ 
The transversal four-dimensional part 
of this metric describes \cite{Townsend} 
the Taub-NUT space. To make it transparent 
we rescale the transversal 
four-dimensional coordinates to 
\begin{eqnarray}  
(\vec{r},\sigma) 
\equiv 
\frac{2}{R}(\vec{x}_{\bot},x^{10}). 
\label{r and sigma}
\end{eqnarray}
With these rescaled coordinates 
the eleven-dimensional metric 
(\ref{D6 solution in M}) becomes 
\begin{eqnarray} 
ds_{11}^2= 
d \vec{x}_{\|}\cdot d \vec{x}_{\|}
+ R^2  \left \{ 
\frac{1}{4}\left(1+\frac{1}{r}\right) 
d \vec{r}\cdot d \vec{r}
+\frac{1}{4\left(1+\frac{1}{r}\right)}
\left(d \sigma
+\vec{\omega}\cdot d\vec{r} 
\right)^2 
\right \}, 
\label{TN metric}
\end{eqnarray} 
where the transversal part acquires  
the standard form \cite{Hawking} 
of the Taub-NUT metric (scaled by $R^2$).

              When there exist 
$N_f$ parallel sixbranes 
their transversal four-dimensions 
becomes the multi-Taub-NUT space. 
The multi-Taub-NUT metric 
acquires the following standard  
form \cite{Hawking} in our coordinates :     
\begin{eqnarray}
ds^2= 
R^2 \left\{
\frac{V(\vec{r})}{4}d \vec{r}\cdot d \vec{r} 
+\frac{1}{4V(\vec{r})}
(d\sigma+\vec{\omega} \cdot d \vec{r})^2 
\right \},
\label{mTN metric}
\end{eqnarray}
The potential $V(\vec{r})$ is 
given by the superposition  
\begin{eqnarray} 
V(\vec{r})=1+ 
\sum_{i=1}^{N_f}\frac{1}{|\vec{r}-\vec{r}_i|}, 
\end{eqnarray} 
where $\vec{r}_i$ denotes 
the position of the $i$-th sixbrane. 
The $U(1)$ gauge field 
$A=\vec{\omega}\cdot d \vec{r}$ is 
determined by the relation 
\begin{eqnarray}
\vec{\nabla}_r \times 
\vec{\omega}=\vec{\nabla}_{r}V(\vec{r}).
\end{eqnarray}

          In order to proceed further 
let us give some comments 
on sixbranes and the Taub-NUT geometry. 
First, 
the positions of sixbranes denoted by $\vec{r}_i$ 
look like singularities 
in the metric (\ref{mTN metric}). 
This is incorrect. 
Actually, 
parallel sixbranes are the removable 
NUT singularities in their transversal 
multi-Taub-NUT space as far as they do not coincide. 
This means that 
{\it sixbranes become geometry in $M$-theory}. 
Second, the multi-Taub-NUT space is known to be 
a hyper-K\"ahler manifold. 
Regarding it as a complex manifold 
$N_f$ parallel sixbranes play 
the role of resolution or deformation 
of the simple singularity of type $A_{N_f-1}$.

\subsubsection{Hyper-K\"ahler structure of the multi-Taub-NUT space}

     We want to consider 
a configuration of $M$-theory fivebrane 
which realize $N\!=\!2$ SQCD with $N_f$ flavors. 
For this purpose 
two-dimensional part $\Sigma$ of the fivebrane 
must be embedded into 
the above multi-Taub-NUT space.  
To discuss its BPS saturation 
we need the hyper-K\"ahler structure 
\cite{NOYY1} \cite{Hitchin}
of the multi-Taub-NUT space.

     Let us introduce   
the complex structure \cite{NOYY1} \cite{Hitchin} 
of the multi-Taub-NUT space $Q$ 
appropriate to describe the BPS saturation 
of the fivebrane.  
For this purpose we need to comment on 
the Dirac strings associated 
with the sixbranes. 
If one regards $Q$ as a Riemannian manifold 
the direction of the Dirac strings is arbitrary. But, 
once one introduces a complex structure, 
their direction becomes unique. 
This is because 
putting a complex structure means restricting 
the structure group from $SO(4)$ to $U(2)$. 
This $U(2)$ simply acts as $SO(2)$ on $\vec{r}$ 
and can not change the direction of the Dirac strings. 
So, taking it reverse, 
in order to introduce a complex structure 
we need to fix the Dirac strings of the sixbranes. 
Let us separate $\vec{r}$ into two parts  
$\vec{r}=(v,b)$. (See eq.(\ref{v b sigma}).) 
Denote the position of the $i$-th sixbrane 
by $\vec{r}_i=(e_i,b_i)$. 
We will fix 
the $i$-th Dirac string as the semi-infinite 
line in the $(v,b)$-space running from 
$\vec{r}_i=(e_i,b_i)$, parallel with the $b$-axis, 
into $(e_i, +\infty)$. 
They are associated with the following 
$U(1)$ gauge field $A=\vec{\omega}\cdot d \vec{r}$ : 
\begin{eqnarray} 
\vec{\omega}\cdot d\vec{r}= 
\mbox{Im}\left( 
\sum_{j=1}^{N_f}\frac{1}{\Delta_j}
\frac{b-b_j+\Delta_j}{v-e_j}~
d v \right), 
\label{gauge fixed A}
\end{eqnarray} 
where 
\begin{eqnarray} 
\Delta_j  & \equiv & 
\sqrt{(b-b_j)^2+ |v-e_j|^2} 
\nonumber \\ 
&=& 
|\vec{r}-\vec{r}_j| .
\label{Delta}
\end{eqnarray} 
The field strength of $A$ becomes 
\begin{eqnarray} 
F \sim \pi i 
\sum_{j=1}^{N_f} 
\theta(b-b_j)\delta^{(2)}(v-e_j)dv 
\wedge \overline{dv}, 
\label{Dirac strings} 
\end{eqnarray} 
which shows that the Dirac strings are fixed 
in the way we expect.

     Now let us introduce the following function 
$y(v,b,\sigma)$ on $Q$ : 
\begin{eqnarray} 
y(v,b,\sigma) 
\equiv 
C e^{-\frac{b+i \sigma}{2}} 
\prod_{j=1}^{N_f}(-b+b_j+\Delta_j)^{\frac{1}{2}} ,
\label{y(v,b,sigma) of mTN}
\end{eqnarray} 
where $C$ is some constant. 
Notice that the differential $dy$ can 
be expressed in terms 
of $dv,db$ and $d \sigma$ : 
\begin{eqnarray} 
\frac{2dy}{y}= 
-V db -i d\sigma + \mbox{Re} 
\left(  \delta~ dv \right), 
\label{dy of mTN} 
\end{eqnarray} 
where we abbreviate the notation  
\begin{eqnarray}
\delta
\equiv  
\sum_{i=1}^{N_f}
\frac{1}{\Delta_i}\frac{b-b_i+\Delta_i}{v-e_i}. 
\end{eqnarray}

                 The multi-Taub-NUT metric 
(\ref{mTN metric}) with the fixed gauge potential 
(\ref{gauge fixed A}) can be rewritten 
only in terms of $v$ and $y$ : 
\begin{eqnarray}
ds^2 
&=& 
R^2 \left[ 
\frac{V}{4}(db^2+dv d \overline{v})
+\frac{1}{4V}\left\{ d\sigma +\mbox{Im}(\delta~dv) \right\}^2 
\right] 
\nonumber \\ 
&=& 
R^2 \left[ 
\frac{V}{4}dv d \overline{v} 
+\frac{1}{4V}
       \left\{ V^2 db^2 
           +\left(d\sigma + \mbox{Im}(\delta~dv)\right)^2 
              \right\} \right] 
\nonumber \\ 
&=& 
R^2 \left[
       \frac{V}{4}dv d \overline{v}
      +\frac{1}{4V}
         \left\{ 
         \left\{ \mbox{Re}
             \left(2dy/y~-\delta dv \right) 
                \right\}^2 
         + 
         \left\{ \mbox{Im}
              \left(2dy/y~-\delta dv \right) 
                 \right\}^2 
        \right\} \right]
\nonumber \\ 
&=& 
R^2 \left\{
\frac{V}{4}dvd\bar{v}
+\frac{1}{4V}\left(\frac{2dy}{y}-\delta dv\right)
\overline{\left(\frac{2dy}{y}-\delta dv\right)} 
\right\}, 
\label{Kahler metric of mTN}
\end{eqnarray}
where we utilize eq.(\ref{dy of mTN}) to derive 
the third equality. 
This shows that 
$y$ in (\ref{y(v,b,sigma) of mTN}) and $v$ give 
the holomorphic coordinates of $Q$ with which  
the metric becomes K\"ahlerian. 
The K\"ahler and holomorphic two-forms are given by 
\begin{eqnarray}
\Omega_{\bf R}
&=&
\frac{i R^2}{2}\left\{ 
\frac{V}{4}dv\wedge d\bar{v}
+\frac{1}{4V}\left(\frac{2dy}{y}-\delta dv\right)\wedge
\overline{\left(\frac{2dy}{y}-\delta dv\right)} 
\right\}, 
\nonumber \\ 
\Omega_{\bf C}
&=& 
\frac{iR^2}{2}dv\wedge\frac{dy}{y}. 
\end{eqnarray}

\subsubsection{Embedding of the Seiberg-Witten curve}

Since the sixbranes play the role of
matter hypermultiplets, the Seiberg-Witten curve 
which is a part of the fivebrane changes \cite{Witten1}
to the curve of $N\!=\!2$ supersymmetric QCD 
with $N_f$ flavors \cite{HO,APSh}
\begin{eqnarray}
y^2-2\prod_{a=1}^{N_c}(v-\phi_a)y
+\Lambda^{2N_c-N_f}\prod_{i=1}^{N_f}(v-e_i)=0,
\label{with matter}
\end{eqnarray}
where $e_i$ are interpreted as the bare masses 
of matter hypermultiplets and 
identified with the positions 
of sixbranes in the $v$-plane.
Now the above hyper-elliptic curve $\Sigma_{SW}$ 
is embedded into the multi Taub-NUT space $Q$. 
Let $\Sigma_{\pm}$ be its two Riemann sheets  
\begin{eqnarray} 
\Sigma_{SW}= \Sigma_{+} \cup \Sigma_{-}. 
\end{eqnarray} 
These Riemann sheets $\Sigma_{\pm}$ are 
realized in $Q$ by the equations 
\begin{eqnarray} 
y_{\pm}(v)=y(v,b,\sigma) ,
\end{eqnarray} 
where $y_{\pm}(v)$ are two roots of eq.(\ref{with matter})
\begin{eqnarray}
y_{\pm}(v)= 
\prod_{a=1}^{N_c}(v-\phi_a)
\pm \sqrt{\prod_{a=1}^{N_c}(v-\phi_a)^2
-\Lambda^{2N_c-N_f}
\prod_{i=1}^{N_f}(v-e_i)}, 
\end{eqnarray}
and the holomorphic coordinate $y$ is regarded as 
the function (\ref{y(v,b,sigma) of mTN}) 
of $v,b$ and $\sigma$.


\section{Minimal area membranes and BPS states}

                     In this section we study 
minimal area membranes which are related with the 
BPS states of $N\!=\!2$ SQCD. 
$M$-theory brane configuration described 
in the previous section is treated as background 
geometry for these membranes.

         Consider a membrane which 
worldvolume is ${\bf R}\times D$, 
where $D$ is a two-dimensional surface 
embedded in the multi-Taub-NUT space $Q$. 
Area of the membrane measured by the induced metric  
satisfies the inequality \cite{FS1,HY,Mikhailov}
\begin{eqnarray}
A_D \geq 
\left| 
\int_D 
\Omega_{\bf{C}} 
\right|. 
\label{BPS inequality}
\end{eqnarray} 
The equality, that is, the BPS saturation 
occurs \cite{HY} 
when $D$ satisfies the following two conditions  
\begin{itemize}
\item $\Omega_{\bf{R}}$ vanishes on $D$. 
\item $\Omega_{\bf{C}}$ has a constant phase on $D$. 
\end{itemize}

                These membranes will have 
their correspondences in Type IIA theory. 
Let $\pi$ be the following natural projection 
\begin{eqnarray}
\pi~~  Q \longrightarrow {\bf R}^3 ~~~~~~~~~~ 
\pi(v,b,\sigma)=(v,b).
\end{eqnarray} 
It is simply a map forgetting 
about the eleven-dimensional 
circle of M-theory.
When $\mbox{dim}_{\bf R}\pi (D)\!= \!1$,
the membrane will give rise to 
a fundamental string  
in Type IIA theory.  
When $\mbox{dim}_{\bf R}\pi (D) \!= \!2$, 
It will give rise to 
a D twobrane. 

\subsection{Fundamental strings} 

                     In this case 
$\mbox{dim}_{\bf R}\pi (D)\!=\!1$ holds. 
Let $D$ be a two-dimensional cylinder. 
The boundary of D consists of two circles. 
$\partial D \!=\! S^1 \cup S^1.$ 
The membrane will be classified 
by its boundary condition 
in $Q$. There are three cases : 
\begin{enumerate} 
\item 
Both two boundary circles of $D$ are on $\Sigma_{SW}$ 
in $Q$ and, regarding them as one-dimensional 
cycles of $\Sigma_{SW}$, are homotopic 
to the $\alpha$-cycles. 
\item 
One boundary circle of $D$ 
is on $\Sigma_{SW}$ in $Q$ and, 
regarding it a one-dimensional 
cycle of $\Sigma_{SW}$, 
is homotopic to the $\alpha$-cycle. 
The other boundary circle is 
attached at the sixbrane. 
\item 
Both two boundary circles of $D$ are 
attached at the sixbranes. 
\end{enumerate}
Membrane satisfying the first boundary condition 
becomes a 4-4 string in Type IIA theory and provides  
a vector multiplet of the worldvolume gauge theory.  
Membrane satisfying the second boundary condition 
becomes a 4-6 string in Type IIA theory and provides 
a hypermultiplet of the worldvolume gauge theory. 
Membrane with the last boundary condition 
becomes a 6-6 string in Type IIA theory 
and is irrelevant in our discussion because 
of the decoupling from the fivebrane.

     Before presenting investigations 
on these boundary conditions 
we examine an implication of the condition,  
$\mbox{dim}_{\bf R}\pi (D)\!=\!1$. 
Let us write down the symplectic forms 
$\Omega_{\bf{R},\bf{C}}$ 
in terms of $v,b$ and $\sigma$ 
using the relation (\ref{dy of mTN}) : 
\begin{eqnarray} 
\Omega_{\bf{R}}
&=& 
\frac{iR^2}{4}\left \{ 
\frac{V}{2}dv \wedge d \bar{v} 
-i d b \wedge 
\left (d \sigma + \mbox{Im} 
\left( \delta d v \right) 
\right ) \right \}, 
\nonumber \\ 
\Omega_{\bf{C}}
&=& 
\frac{iR^2}{4} 
d v \wedge 
\left( \frac{\bar{\delta}}{2} 
d \bar{v}- V d b -i d \sigma \right) .
\label{symplectic forms} 
\end{eqnarray} 
Since $\pi (D)$ is a one-dimensional object 
in the ${\bf R}^3$, 
$d v \wedge d \bar{v}$ and $d b \wedge d v$ 
will vanish on $D$. It means that 
the symplectic forms restricted on D 
have the forms 
\begin{eqnarray} 
\left. \Omega_{\bf{R}} \right |_{D} 
=\frac{R^2}{4} d b \wedge d \sigma ~~~~,~~~~~
\left. \Omega_{\bf{C}} \right |_{D} 
=\frac{R^2}{4} d v \wedge d \sigma.  
\label{symplectic forms on D} 
\end{eqnarray}
While this, 
since $D$ is a two-dimensional cylinder, 
$d \sigma $ can not vanish on $D$. 
So, the vanishing of $\Omega_{\bf{R}}$ on $D$, 
which one needs for the BPS saturation, 
requires that 
{\it $D$ is included in the hypersurface of $Q$ 
characterized by $b\!=\! const.$}

\subsubsection{4-4 strings} 

             Let us consider a membrane which satisfies  
the first boundary condition. 
Namely $\partial D$ is on $\Sigma_{SW}$ in $Q$ and  
are homotopic to the $\alpha$-cycles in $\Sigma_{SW}$. 
Since $\pi (D)$ is one-dimensional, 
the projection of each boundary circle of $D$ 
can be expected to be a point on $\pi (\Sigma_{SW})$ 
or a one-dimensional line on $\pi (\Sigma_{SW})$. 
What actually occurs is the latter. 
To explain this, let us recall 
two Riemann sheets $\Sigma_{\pm}$ of $\Sigma_{SW}$ 
are patched together in $Q$ by $N_c$ circles $C_a$ :  
\begin{eqnarray}
\Sigma_+ \cap \Sigma_- 
= C_1 \cup \cdots \cup C_{N_c} ~~~~~
(\subset Q). 
\label{Ca} 
\end{eqnarray} 
These circles, if one regards them as 
one-dimensinal cycles of $\Sigma_{SW}$,  
are homotopic to the $\alpha$-cycles. 
Notice that they are projected by $\pi$ 
to one-dimensional lines $L_a$ in the ${\bf R}^3$ 
\begin{eqnarray} 
L_a = \pi (C_a)  ~~~~~~ (\subset {\bf R}^3),    
\label{La}
\end{eqnarray} 
which are the intersection of $\pi(\Sigma_{\pm})$. 
Taking account of $\pi (D)$ being one-dimensional, 
the boundary circles of D must be attached to $C_a$ 
\begin{eqnarray}
\partial D = C_a \cup C_b. 
\label{boundary condition 1}
\end{eqnarray} 
In particular they satisfy 
\begin{eqnarray} 
\pi(\partial D)= L_a \cup L_b.  
\end{eqnarray}

   On the other hand 
the vanishing of $\Omega_{\bf{R}}$ on $D$ 
requires that 
$D$ must lie in the hypersurface $b \!=\! const.$. 
Combining this requirement with 
eq.(\ref{boundary condition 1}) 
we can conclude that all 
$C_a$ must lie in this hypersurface 
for the vanishing of $\Omega_{\bf{R}}$ on $D$. 
Notice that this is not the 
condition on the membrane but the condition 
on the background geometry .
It can be fulfilled by setting all 
the $b$-positions of the sixbranes equal. 
Let us specialize them to 
\begin{eqnarray} 
b_1 = \cdots = b_{N_f}=0 .
\label{condition on D6}
\end{eqnarray} 
In this background geometry 
all $C_a$ lie in the $b\!=\!0$ hypersurface of Q. 
One may suspect condition (\ref{condition on D6}) 
loses a generality. But, 
from the viewpoint of the worldvolume gauge theory, 
the brane configuration under consideration 
describes the Coulomb branch and 
the $b$-direction is irrelevant 
in the description of this branch.

     Now $\pi(D)$ is a one-dimensional line in 
the $b\!=\!0$ plane (which we will identify with 
the $v$-plane). It has the form 
\begin{eqnarray} 
\pi(D)= L_a \cup L_{ab} \cup L_b
\label{image of 4-4 string}
\end{eqnarray} 
where $L_{ab}$ is a one-dimensional line 
connecting $L_a$ and $L_b$. 
Let us give somewhat an explict description 
of $L_a$. 
Notice that the intersection of $\pi(\Sigma_{\pm})$ 
can be realized as the solution of the equation 
\begin{eqnarray} 
|y_+(v)| = |y_-(v)|,    
\end{eqnarray}
This equation tells that it 
is the region of the $v$-plane on which  
$\frac{\prod_a(v-\phi_a)^2}
{\Lambda^{2N_c-N_f}\prod_j(v-e_j)}$ 
becomes a real non-negative quantity 
not bigger than $1$. 
It consists of $N_c$ one-dimensional lines 
connecting 
$v\!=\! \phi_a^{(+)}$ and 
$v\!=\! \phi_a^{(-)}$ through $v\!=\! \phi_a$. 
Here $\phi_a^{(\pm)}$ are 
the branch points of the curve 
which approach to $\phi_a$ 
as $\Lambda \! \rightarrow \! 0$.
These one-dimensional lines are $L_a$.

            Based on the form of $\pi(D)$ given 
in (\ref{image of 4-4 string}), 
we can say $D$ has the following form in $Q$:  
\begin{eqnarray} 
D = D_a \cup (L_{ab} \times S^1) \cup D_b, 
\label{4-4 string}
\end{eqnarray}
where the $S^1$ is the circle 
of the eleven-dimension.    
$D_{a}$ is the region of the cylinder 
$L_{a} \times S^1$ and 
is projected to $L_{a}$ by $\pi$. 
The boundary of $D_{a}$ 
consists of two circles 
one of which is identified with 
the boundary circle of the cylinder 
$L_{ab}\times S^1$ 
and the other is identified with 
the $\alpha$-cycle $C_{a}$. 
Recall $C_a$ is the circle 
wrapping once $L_{a} \times S^1$ 
and is projected to $L_{a}$ by $\pi$.  
$D_{a}$ occupies the half of $L_a \times S^1$ 
separated by $C_{a}$.

       The area of the membrane is the sum of those of 
$D_{a,b}$ and $L_{ab}\times S^1$. The area of $D_{a}$ is 
half the area of $L_{a} \times S^1$. Therefore it holds 
\begin{eqnarray} 
A_D= 
A_{L_{ab}\times S^1}+ 
\frac{1}{2}(A_{L_a \times S^1}+ A_{L_b \times S^1}).  
\label{area of 4-4 string} 
\end{eqnarray}   
Let $L \times S^1$ be a cylinder in $Q$ realized by 
$v\!=\!v(s), b\!=\!0$ and $\sigma \!= \! \sigma(t)$. 
$s$ and $t$ parametrize this cylinder. 
The induced metric $h_{ij}$ of $L \times S^1$ becomes  
\begin{eqnarray} 
h_{ss}&=& 
\frac{R^2}{4}
\left \{ V \left|\frac{dv}{ds}\right|^2
+\frac{1}{V} 
\left(\mbox{Im}
\left(\delta \frac{dv}{ds}\right)
\right)^2 
\right \}, 
\nonumber \\ 
h_{st}&=& 
\frac{R^2}{4} 
\frac{1}{V}
\mbox{Im}
\left(\delta \frac{dv}{ds}\right) 
\frac{d \sigma}{dt},  
~~~~~~
h_{tt}= 
\frac{R^4}{4}\frac{1}{V}
\left(\frac{d \sigma}{dt}\right)^2 .
\end{eqnarray} 
The volume form  
determined by this metric is 
\begin{eqnarray} 
d vol_{L \times S^1}  
&=& \frac{R^2}{4} \left|\frac{dv}{ds}\right| 
\frac{d \sigma}{dt} ds \wedge dt  
\nonumber \\ 
&\equiv& \frac{R^2}{4} |dv(s)| \wedge d \sigma(t). 
\end{eqnarray}
Therefore the area of $L \times S^1$ is 
$\pi R^2 |L|$. ($|L| \equiv \int_{L}|dv|$.) 
Now we can evaluate eq.(\ref{area of 4-4 string}) to  
\begin{eqnarray} 
A_D=
\pi R^2 
\left(|L_{ab}|+\frac{|L_a|+|L_b|}{2}\right). 
\label{area of 4-4 string 2}
\end{eqnarray}

               At this stage 
we want to minimize this $A_D$. Notice that 
$L_{a}$ is fixed by the background geometry 
\footnote{$M$-theory brane configurations 
described in the previous section 
are treated as background geometry for membranes}. 
Only parameter which we can change is $L_{ab}$. 
It is the line connecting $L_{a}$ and $L_{b}$. 
Clearly we must take the straight line 
for the minimization. 
The minimal area is 
\begin{eqnarray} 
A_{min}=
\pi R^2 
\left(|\phi_a^{(-)}-\phi_b^{(+)}|
+ \frac{|L_a|+|L_b|}{2}\right). 
\label{minimal area of 4-4 string} 
\end{eqnarray}  
But 
{\it this minimal area surface does not 
saturate the BPS bound.}  This is because, 
though $\Omega_{\bf{R}}$ vanishes on $D$ by 
the construction, 
the phase of $\Omega_{C}$ changes on $D_{a,b}$  
and we can not tune it 
to the definite phase which  
$\Omega_{\bf{C}}$ takes on $L_{ab}\times S^1$. 
The parts $D_{a,b}$ of the membrane are fixed 
by the background geometry. 
To achieve the BPS bound we need to change 
the background to a specific one 
by adjusting the values of 
the moduli parameters of the brane configuration. 
This will be discussed in the later subsection. 
But from the viewpoint describing the Coulomb branch,  
it is not favorable 
because of the loss of generality.  
Actually there exists another way 
to make the BPS saturated membrane. 
It is to consider a possibility of some recoil of 
the membrane to the background geometry. 
But to pursue such a possibility 
is beyond this paper.

              The BPS bound has 
a nice interpretation \cite{FS1,HY,Mikhailov}
in terms of the Seiberg-Witten differential 
$\lambda_{SW} = v \frac{dy}{y}$ on the curve : 
\begin{eqnarray}
\left|\int_D \Omega_{\bf C}\right| 
= \frac{R^2}{2}\left|\oint_{Ca-Cb} \lambda_{SW}\right|. 
\end{eqnarray} 
Let us estimate the deviation of the minimal area  
from the BPS bound in the region 
$|\phi_a \!-\! \phi_b|,|\phi_a \!-\!e_j| >\!\!> \Lambda$ .
One can estimate $|L_{a}|$ as 
$|L_{a}| \! \approx \! \Lambda$  in this region. 
Therefore the deviation is approximately 
\begin{eqnarray}
\frac{A_{min}-\left|\int_D \Omega_{\bf{C}}\right|}
{\left|\int_D \Omega_{\bf{C}}\right|} 
\approx 
\frac{\Lambda}{|\phi_a-\phi_b|} 
\end{eqnarray} 
One can also discuss 
the mass of the minimal area membrane 
as an expansion by $\Lambda /|\phi_a\!-\!\phi_b|$. 
It has the form 
\begin{eqnarray} 
\frac{A_{min}}{l_p^3}
=\frac{\pi R^2}{l_p^3}|\phi_a-\phi_b|
\left(1+ O \left(\frac{\Lambda}{|\phi_a-\phi_b|} 
\right)\right). 
\end{eqnarray} 
Rewriting the leading term 
by the original ten-dimensional coordinates 
it can be read as  
$\frac{\pi}{l_s^2}
|(x_a^4\!+\!ix_a^5)\!-\!(x_b^4\!+\!ix_b^5)|$,  
where $(x_a^4,x_a^5)$ are 
the positions of the fourbranes 
in the $x^{4,5}$-directions.  
This is an expected result from Type IIA theory. 

\subsubsection{4-6 strings}

            We will consider a membrane 
which satisfies 
the second boundary condition. 
The background geometry is same as 
in the case of 4-4 strings. 
That is, $b_1= \cdots = b_{N_f}\!=0$. 
Two-dimensional part $D$ of the membrane 
is a cylinder which two boundary  
circles are attached respectively 
to the $\alpha$-cycle $C_a$ and 
the $j$-th sixbrane. 
For the vanishing of $\Omega_{\bf R}$ 
the cylinder $D$ must lie 
in the hypersurface $b\!=\!0$. 
Following the notation of the previous study  
we can conclude 
$D$ must have the following form in $Q$ :
\begin{eqnarray} 
D = (L_{ja}\times S^1) \cup D_a  ~~~~~( \subset Q),  
\label{4-6 string} 
\end{eqnarray}
where 
$L_{ja}$ is a one-dimensional 
line in the $v$-plane 
(which we identify with the $b\!=\!0$ plane 
of the ${\bf R}^3$) 
connecting $v\!=\!e_j$ and $L_a$.  
The area of $D$ becomes   
\begin{eqnarray} 
A_D= 
\pi R^2 
\left( |L_{ja}| + \frac{|L_a|}{2} \right).  
\end{eqnarray}
One can minimize the area by making 
$L_{ja}$ the straight line. 
Again, notice that $L_a$ is fixed 
by the background geometry. 
The minimal area is  
\begin{eqnarray} 
A_{min}= 
\pi R^2 \left( 
|e_j-\phi_a^{(+)}|+\frac{|L_a|}{2} 
\right).
\label{minimal area of 4-6 string}
\end{eqnarray}

            But, this minimal area surface 
does not saturate the BPS bound by 
the same reason as in the case of 
4-4 strings. 
The BPS bound has a nice interpretation 
in terms of $\lambda_{SW}$ 
besides the bare mass 
\begin{eqnarray} 
\left |\int_D \Omega_{\bf C} \right |= 
\frac{R^2}{2}
\left |2 \pi  e_j- 
\oint_{C_a} \lambda_{SW} \right | . 
\end{eqnarray}  
The deviation of the minimal area from the 
BPS bound can be estimated in the region 
$|\phi_a\!-\! \phi_b|,|\phi_a\!-\!e_j|>\!\!> \Lambda$
\begin{eqnarray}
\frac{A_{min}-\left|\int_D \Omega_{\bf{C}}\right|}
{\left|\int_D \Omega_{\bf{C}}\right|} 
\approx 
\frac{\Lambda}{|e_j-\phi_a|}.  
\end{eqnarray} 
In this region the mass of the minimal area membrane 
has the form 
\begin{eqnarray} 
\frac{A_{min}}{l_p^3}=
\frac{\pi R^2}{l_p^3}|e_j-\phi_a|
\left(1+ O \left(\frac{\Lambda}{|e_j-\phi_a|} 
\right)\right). 
\end{eqnarray}
Rewriting the leading part in terms of 
the original ten-dimensional coordinates 
it can be found to give 
an expected result from Type IIA theory. 

\subsubsection{6-6 strings}

            We will consider a membrane 
which satisfies the third boundary condition. 
The background geometry is same as 
in the previous cases. 
Two-dimensional part $D$ of the membrane 
is a cylinder which two boundary circles are 
attached respectively 
to the $j$-th and $k$-th sixbranes. 
For the vanishing of $\Omega_{\bf R}$ 
the cylinder $D$ must lie 
in the hypersurface $b\!=\!0$. 
Following the same notation as before  
$D$ must have the following form 
in $Q$ :  
\begin{eqnarray} 
D = L_{jk}\times S^1  ~~~~~~~( \subset Q),  
\label{6-6 string} 
\end{eqnarray}
where $L_{jk}$ is a one-dimensional 
line in the $v$-plane 
connecting $e_j$ and $e_k$ .  
The area of $D$ becomes   
\begin{eqnarray} 
A_D= \pi R^2 |L_{jk}| .
\end{eqnarray}
One can minimize the area by making 
$L_{jk}$ the straight line.   
The minimal area is 
\begin{eqnarray} 
A_{min}= \pi R^2 |e_j-e_k| .
\label{minimal area of 6-6 string}
\end{eqnarray}
This minimal area surface saturate the BPS bound 
since $\Omega_{\bf C}$ has a constant phase on the 
straight line $L_{jk}$.


\subsection{Dirichlet two-branes}

        In this case $\mbox{dim}_{\bf R} D=2$ holds. 
Let $D$ be a two-dimensional disk. 
We can impose the following 
boundary condition on $D$ 
\begin{itemize}
\item 
The boundary circle of $D$ is attached on 
$\Sigma_{SW}$ in $Q$ 
and, considering it as a one-dimensional cycle of 
$\Sigma_{SW}$, is homotopic to the $\beta$-cycle. 
\end{itemize} 
Membrane satisfying this boundary condition will 
gives rise to a D twobranes and provides a monopole 
in the worldvolume gauge theory.

        This boundary condition is so severe 
in a general background geometry 
that we do not know how to minimize the area  
preserving it. 
The solution we have is very restrictive. 
Let us specialize \cite{HY} 
the background geometry to 
\begin{eqnarray} 
\forall \phi_a,~~ \forall e_j \in {\bf R} . 
\end{eqnarray} 
Notice that the previous minimal area membranes 
corresponding to 4-4 and 4-6 strings 
also saturate their BPS bounds in this 
specific background geometry. 
This is simply because 
all $L_a$ lie in the real line of the $v$-plane.  
In this background geometry 
one can take representatives 
of the $\beta$-cycles such that they are  
one-dimensional circles 
within the hypersurface $v=\bar{v}$ 
of $Q$. Let us denote them by 
$B_a$ ($1 \leq a \leq N_c-1$). 
One can easily see that 
$\sigma$ takes constant values 
on these circles. 
Let $D$ be a disk in the hypersurface 
$v=\bar{v}$ with its boundary being $B_a$ and 
its $\sigma$-position being same as the boundary. 
Now on this disk 
$\Omega_{\bf R}$ vanishes and $\Omega_{\bf C}$ 
has a constant phase. 
So this membrane saturates the BPS bound 
as follows : 
\begin{eqnarray} 
A_D &=& ~~~~~
\left| \int_{D} \Omega_{\bf C} \right| 
\nonumber \\ 
&=& 
\frac{R^2}{2} 
\left| \oint_{B_a} \lambda_{SW} \right|. 
\end{eqnarray}


\section{Higgs Branch of $N\!=\!2$ MQCD}

                In this section 
we study the Higgs branch of $N\!=\!2$ MQCD. 
We first review the field theory results 
mainly based on \cite{APS}. 
Consider $N\!=\!2$ $SU(N_c)$ supersymmetric QCD 
with $N_f$ flavors. We will treat the case of 
$N_f \! \leq \! N_c \!<\! 2N_f $. 
The classical vacua of this theory 
described by the $F$- and $D$-flat conditions. 
Let us denote the hypermultiplets by 
$(Q,\tilde{Q})$. 
$N\!=\!1$ chiral multiplets $Q$ and $\tilde{Q}$ 
respectively belong to 
the fundamental and anti-fundamental 
representations of the gauge group $SU(N_c)$ 
besides the flavor group $U(N_f)$. 
We realize $Q$ as a complex 
$N_c \! \times \! N_f$ matrix and 
$\tilde{Q}$ as a complex 
$N_f \! \times \! N_c$ matrix.   
In the Higgs branch 
the $F$- and $D$-flat conditions become 
\begin{eqnarray} 
Q \tilde{Q} &=& \zeta_{\bf C}{\bf 1}_{N_c}, 
\nonumber \\ 
Q Q^{\dagger}-\tilde{Q}^{\dagger}\tilde{Q}
&=& \zeta_{\bf R}{\bf 1}_{N_c}  ,
\label{flat conditions}
\end{eqnarray} 
where $\zeta_{{\bf C},{\bf R}}$ are respectively 
complex and real numbers which we will call 
the Fayet-Iliopoulos (FI) parameters. 
$\vec{\zeta} \! 
\equiv \!(\zeta_{\bf C},\zeta_{\bf R})$. 
Consideration on the Higgs branch is separated 
into two cases whether the FI parameters 
vanish or not. 
The former case is non-baryonic and 
the latter is baryonic. 
In the baryonic branch 
the color symmetry is completely 
broken due to the non-vanishing FI parameters. 
In the non-baryonic branch 
some of the color symmetry 
remains unbroken.

      Let us examine the baryonic branch first. 
Fix the non-vanishing FI parameters at some values. 
Introducing the functions  
$\mu_{\bf C}(Q,\tilde{Q})$
$\equiv Q\tilde{Q}$ 
and 
$\mu_{\bf R}(Q,\tilde{Q})$
$\equiv QQ^{\dagger} \!-\! 
\tilde{Q}^{\dagger}\tilde{Q}$,    
we can write the moduli space of 
the (gauge equivalent) solutions 
of eqs.(\ref{flat conditions}) as 
\begin{eqnarray} 
\frac{\mu_{\bf C}^{-1}(\zeta_{\bf C}) \cap 
              \mu_{\bf R}^{-1}(\zeta_{\bf R})}
     {SU(N_c)} 
&\simeq&  
U(1) \times 
\frac{\mu_{\bf C}^{-1}(\zeta_{\bf C}) \cap 
              \mu_{\bf R}^{-1}(\zeta_{\bf R})}
     {U(N_c)} , 
\nonumber \\ 
&\equiv& 
U(1)\times 
\hat{{\cal M}}_{\mbox{b}}(\vec{\zeta}),
\end{eqnarray} 
where we rewrite the LHS using the isomorphism 
$U(N_c)\! \simeq \!U(1) \times SU(N_c)$. 
The second factor in this factorization, 
that is, the space 
$\hat{{\cal M}}_{\mbox{b}}(\vec{\xi})$ 
is a hyper-k\"ahler manifold. 
With a little calculation 
one can find out 
it is a $4N_c(N_f-N_c)$-dimensional 
smooth manifold. 
The FI parameters $\vec{\zeta}$ 
besides the $U(1)$ appearing 
in the above factorization can be also 
regarded as parameters of the baryonic branch. 
Let us denote their moduli space by $P$. 
Then the moduli space of the baryonic branch, 
which we denote by ${\cal M}_{\mbox{b}}$, 
is the fibre bundle over $P$. 
The fibre space 
at $(\vec{\zeta},e^{i \phi}) \in P$ is 
$\hat{{\cal M}}_{\mbox{b}}(\vec{\zeta})$. 
So, ${\cal M}_{\mbox{b}}$ is the moduli space of 
$N_c(N_f-N_c)+1$ massless hypermultiplets.

                   Next, let us consider 
the non-baryonic branch. 
It is characterized by 
the vanishing FI parameters. 
The moduli space of the non-baryonic branch, 
which we denote by 
${\cal M}_{\mbox{nb}}$, have the form : 
\begin{eqnarray} 
{\cal M}_{\mbox{nb}} \equiv 
\frac{\mu_{\bf C}^{-1}(0) 
\cap \mu_{\bf R}^{-1}(0)}{U(N_c)}. 
\end{eqnarray} 
This is a singular manifold but it is possible to give a 
stratification by the color symmetry 
breaking patterns. 
For the breaking pattern, 
$SU(N_c)\! \rightarrow \!SU(N_c-r)$ 
we can introduce a stratum 
${\cal M}_{\mbox{nb}}^{(r)}$ 
as the space of the (gauge equivalent) solutions 
of $\mu_{{\bf C}}\!=\! \mu_{\bf R}\!=\!0$ 
which break only $U(r)$ color symmetry : 
\begin{eqnarray} 
{\cal M}_{\mbox{nb}} 
= \cup_{r}{\cal M}_{\mbox{nb}}^{(r)}. 
\end{eqnarray} 
Each component ${\cal M}_{\mbox{nb}}^{(r)}$ 
can be considered 
as a hyper-K\"ahler manifold of 
dimensions $4r(N_f-r)$ and 
is called the moduli space of 
the $r$-th non-baryonic branch. 
So, this moduli space includes 
$r(N_f-r)$ massless hypermultiplets.

~

        The Higgs branch intersects 
with the Coulomb branch. 
Intersection of the baryonic branch and 
the Coulomb branch is a 
single point of ${\cal M}_{\mbox{coul}}$, 
the moduli space of the Coulomb branch.  
This is called the baryonic branch root. 
Though $U(1)^{ N_c-1}$ color symmetry 
remains unbroken at a generic 
point of the Coulomb branch, 
it must be completely broken at this root. 
This is due to the requirement 
from the baryonic branch side. 
The intersection of the $r$-th non-baryonic branch 
and the Coulomb branch is 
a $2(N_c-r-1)$-dimensional submanifold 
of ${\cal M}_{\mbox{coul}}$. 
This is called the $r$-th non-baryonic branch root. 
On this submanifold 
only $U(1)^{ N_c-r-1}$ color symmetry 
remains unbroken. 
This is because $U(r)$ color symmetry 
must be broken by the requirement 
from the $r$-the non-baryonic branch side and 
the remaining $SU(N_c-r)$ color symmetry 
is broken down to $U(1)^{ N_c-r-1}$ 
of the low energy effective theory.

~

          In the following subsections 
we present the descriptions of 
the baryonic and non-barynic branches 
in terms of the $M$-theory brane 
configurations. 
We follow the argument given in \cite{NOYY1}. 
We begin by studying their roots using   
the $M$-theory description 
of the Coulomb branch 
in the previous sections. 
In that description, 
the Seiberg-Witten curve, that is, 
two-dimensional part of the $M$ fivebrane, 
is holomorphically embedded 
into the multi-Taub-NUT space $Q$, 
which is the geometry sixbranes 
provide in $M$-theory.  
Moduli parameters of this configuration 
are $\phi_{a} \in {\bf C}$ for the fivebrane 
and $(e_i,b_i) \in {\bf C} \times {\bf R}$ 
for the sixbranes 
(or the multi-Taub-NUT space).    
After specifying these moduli parameters 
at the values relevant for the roots, 
we will show that 
the baryonic and non-baryonic 
branches are actually realized in $M$-theory.

\subsection{Non-baryonic Branch}

       We consider the $r$-th non-baryonic branch 
root, that is the intersection 
of ${\cal M}_{\mbox{nb}}^{(r)}$ 
and ${\cal M}_{\mbox{coul}}$, 
from the Coulomb branch side. 
The corresponding brane configuration 
is given by the conditions,   
\begin{eqnarray} 
\phi_1= \cdots = \phi_r=0 ,~~~~
\phi_a \neq 0 ~~~(r+1 \leq a \leq N_c ) 
\label{M5 for NB root}
\end{eqnarray}
for the fivebrane and 
\begin{eqnarray} 
e_1= \cdots = e_{N_f}=0 
\label{D6 for root 1} 
\end{eqnarray}
for the sixbranes. 
We set the $b$-positions of the sixbranes 
as follows :
\begin{eqnarray} 
b_1 < \cdots < b_{N_f}.  
\label{D6 for root 2}
\end{eqnarray} 
If one takes Type IIA picture 
it will correspond to 
the configuration in which  
$r$ of $N_c$ D fourbranes overlap 
and $N_f$ D sixbranes cross 
these $r$ D fourbranes. 
Their crossing breaks the $U(r)$ 
color symmetry completely.

     To study this configuration 
let us recall compact two-cycles of Q. 
In general there are $N_f-1$ topologically 
nontrivial two-spheres in Q. 
Let us denote them by $S_j$. 
By the natural projection 
$\pi$ $Q \rightarrow {\bf R}^3$, 
each two-sphere $S_j$ is projected 
to the one-dimensional straight 
line $l_j$ connecting 
$(e_j,b_j)$ and $(e_{j+1},b_{j+1})$. 
The symplectic two-forms 
$\Omega_{\bf{R},\bf{C}}$ take the following 
form on $S_j$ : 
\begin{eqnarray} 
\left. \Omega_{\bf{R}}\right |_{Sj}= 
\frac{R^2}{4}db \wedge d \sigma, ~~~~~~~ 
\left. \Omega_{\bf{C}}\right |_{S_j}= 
\frac{R^2}{4} dv \wedge d \sigma . 
\end{eqnarray} 
In particular,  
at the configuration relevant for  
the non-baryonic branch root, 
the holomorphic two-form restricted on $S_j$ 
vanishes. 
This means that all $S_j$ 
are holomorphic in $Q$ at the root.

     At the non-baryonic branch root 
one can expect that 
$\Sigma_{SW}$ wraps these two-cycles $S_j$. 
Wrapping and unwrapping parts of $\Sigma_{SW}$ 
intersect only at the points where the sixbranes 
are located. 
We can regard these wrapping parts of 
the fivebrane as fivebranes  
with worldvolume ${\bf R}^4 \times S_j$.  
They can move 
to the $x^{7,8,9}$-directions on the sixbranes 
without any obstruction. 
Rotation group of these directions 
is identified with $SU(2)_R$ of 
$N\!=\!2$ supersymmetry. 
So, they will provide \cite{Witten1} 
massless hypermultiplets 
in the non-baryonic branch.

     How many massless hypermultiplets 
can we obtain? 
To count their number and 
to check that $\Sigma_{SW}$ 
actually wraps these two-cycles, 
it is convenient to study 
the holomorphic embedding of the curve 
in a different fashion,  
\begin{eqnarray} 
y(v,b,\sigma)+z(v,b,\sigma)
=2v^r \prod_{a=r+1}^{N_c}(v-\phi_a),  
\label{embedding equation} 
\end{eqnarray} 
where 
$z(v,b,\sigma)
\equiv\Lambda^{2N_c-N_f}v^{N_f}/y(v,b,\sigma)$.

         To see whether $\Sigma_{SW}$ 
wraps $S_j$ or not, 
it is enough to show the following : 
For any $b^*$ satisfying 
$b_j \!<\! b^* \!<\! b_{j+1}$, 
$v\!=\!0$ is a solution of 
the embedding equation at $b^*$, 
\begin{eqnarray} 
y(v,b^*,\sigma)+z(v,b^*,\sigma)
=2v^r \prod_{a =r+1}^{N_c}(v-\phi_a). 
\label{embedding equation at b*}
\end{eqnarray} 
Let us explain why this is enough. 
Suppose $v\!=\!0$ is a solution 
of eq.(\ref{embedding equation at b*}) 
for any $b^*$. 
Then $l_j$ ($\!=\! \pi(S_j)$) is 
included in $\pi(\Sigma_{SW})$. 
From the dimensionality 
this means that $S_j$ is 
included in $\Sigma_{SW}$. 
Taking this viewpoint 
the multiplicity of solution $v\!=\!0$ 
for eq.(\ref{embedding equation at b*}) 
is nothing but 
the wrapping number 
of $\Sigma_{SW}$ to $S_j$. 
This is the number of the 
fivebranes with worldvolume 
${\bf R}^4 \times S_j$.

              Let us show the above claim. 
Consider eq.(\ref{embedding equation at b*}) 
in the region 
$|v| \!<\!<\! |b^*\!-\!b_j|, |b^*\!-\!b_{j+1}|$. 
In this region $y(v,b^*,\sigma)$ and $z(v,b^*,\sigma)$ 
behave as 
\begin{eqnarray} 
y(v,b^*,\sigma) &\sim& 
\prod_{i=1}^{N_f}
(-b^*+b_i+|b^*-b_i|+\frac{1}{2}|v|^2/|b^*-b_i|)^{1/2}, 
\nonumber \\
z(v,b^*,\sigma) &\sim& 
\prod_{i=1}^{N_f}
(b^*-b_i+|b^*-b_i|+\frac{1}{2}|v|^2/|b^*-b_i|)^{1/2}. 
\label{estimate of y and z}
\end{eqnarray}
Notice that the following equalities 
\begin{eqnarray}
-b+b_i+|b-b_i| &=&
\left\{
\begin{array}{lcl}
2(-b+b_i)~~~~(\neq 0) & \mbox{if} & b<b_i \\
0         & \mbox{if} & b>b_i
\end{array}
\right.,
\nonumber \\
b-b_i+|b-b_i| &=&
\left\{
\begin{array}{lcl}
0 & ~~\mbox{if} & b<b_i \\
2(b-b_i)~~~~( \neq 0)  & ~~\mbox{if} & b>b_i
\end{array}
\right. .
\nonumber 
\end{eqnarray}
So, eqs.(\ref{estimate of y and z}) imply
\begin{eqnarray} 
y(v,b^*,\sigma) &\sim&  
v^j \times 
( \mbox{non-vanishing factor at}~ v=0),  
\nonumber \\ 
z(v,b^*,\sigma) 
&\sim&  
v^{N_f-j} 
\times 
(\mbox{non-vanishing factor at}~ v=0) .
\end{eqnarray} 
Now eq.(\ref{embedding equation at b*}) 
acquires the following form in this region 
\begin{eqnarray} 
v^j + v^{N_f-j} = v^r. 
\label{proof of s-rule}
\end{eqnarray} 
This means that $v\!=\!0$ is 
a solution of 
eq.(\ref{embedding equation at b*}) 
for $b_j \!<\! ^{\forall}b^* \!<\! b_{j+1}$. 
Its multiplicity, which we denote by 
$n_j$, can be read from 
eq.(\ref{proof of s-rule}) 
\begin{eqnarray} 
n_j=
\mbox{min}\left \{ j,N_f-j,r \right \}.  
\label{nj}
\end{eqnarray}
This coincides with the result derived 
in Type IIA picture using the $s$-rule 
\cite{HW}. 
Actually the above argument gives 
a proof of the $s$-rule via M-theory. 
Since all $S_j$ are holomorphic at the root 
one can also show \cite{HOO} 
the $s$-rule as the resolution of the curve.  
Now the total number 
of the massless hypermultiplets 
is the sum of $n_j$  
\begin{eqnarray} 
\sum_{j=1}^{N_f-1}n_j=r(N_f-r) . 
\end{eqnarray} 
This is the correct number 
of massless hypermultiplets 
allowed in the $r$-th non-baryonic branch.

\subsection{Baryonic Branch}

                   The baryonic branch root, 
that is, the intersection 
of ${\cal M}_{\mbox{b}}$ and ${\cal M}_{\mbox{coul}}$, 
is a single point of 
${\cal M}_{\mbox{coul}}$ where the color symmetry 
$SU(N_c)$ is completely 
broken. At a first glance 
one may expect it is given by 
$~^{\forall}\phi_a\!=\!^{\forall}e_j\!=\!0$. 
But this naive expectation 
is incorrect in quantum theory. 
Due to the quantum correction on the Coulomb branch 
the root changes from the naive one. 
Where is the root? 
We first claim that 
the curve is completely degenerate at the root. 
This is because $SU(N_c)$ must be broken at the root. 
In quantum theory  
$U(1)_R$ symmetry is broken 
to ${\bf Z}_{2(2N_c-N_f)}$ by the 
chiral anomaly, 
which acts on the $v$-plane as ${\bf Z}_{2N_c-N_f}$. 
So, we also claim that the curve at the root 
is invariant under this discrete symmetry. 
With these two claims the baryonic branch root 
was determined \cite{APS}. 
It is given by 
\begin{eqnarray} 
&& 
\phi_1= \cdots =\phi_{N_f-N_c}=0,~~~~
\phi_{r}=\omega^{r-N_f+N_c}\Lambda ~~~~~
(N_f-N_c+1 \leq r \leq N_c) ,
\nonumber \\ 
&& 
e_1 = \cdots = e_{N_f}=0 ,
\label{BB root} 
\end{eqnarray} 
where 
$\omega \equiv e^{\frac{2 \pi i}{2N_f-N_c}}$. 
The curve at the root becomes 
\begin{eqnarray} 
y^2-2v^{N_f-N_c}
(v^{2N_c-N_f}-\Lambda^{2N_c-N_f})y 
-4\Lambda^{2N_c-N_f}v^{N_f}=0 .
\label{BB root curve 1}
\end{eqnarray} 
It is invariant under 
the ${\bf Z}_{2N_c-N_f}$ symmetry 
and is completely factorized :
\begin{eqnarray} 
y_+(v)=2v^{N_c},~~~~~
y_-(v)=-2 \Lambda^{2N_c-N_f}v^{N_f-N_c}.
\label{BB root curve 2}
\end{eqnarray}

                  The previous consideration 
on the wrapping of $\Sigma_{SW}$ 
to the two-spheres $S_j$ works also 
on this case. 
Setting $r$ in (\ref{nj}) 
to be $N_f\!-\!N_c$, 
the wrapping number $n_j$ of 
$\Sigma_{SW}$ to $S_j$ is  
\begin{eqnarray} 
n_j= \mbox{min} 
\left \{ j,N_f-j,N_f-N_c \right \}. 
\end{eqnarray} 
So, the total number 
of the massless hypermultiplets 
given by the fivebranes 
with worldvolume ${\bf R}^4 \times S_j$ is 
\begin{eqnarray} 
\sum_{j=1}^{N_f-1}n_j = N_c(N_f-N_c) . 
\end{eqnarray} 
This is the correct number 
of massless hypermultiplets 
in $\hat{{\cal M}}_{\mbox{b}}(\vec{\xi})$.


\section{Duality of Baryonic Branch}

       The baryonic branch root is located 
in the strong coupling region of the original 
microscopic $SU(N_c)$ gauge theory. 
Gauge theory description in the neighborhood of 
the root can be obtained by applying 
electric-magnetic duality. 
The magnetic theory is \cite{APS}
the $SU(N_f\!-\!N_c)\times U(1)^{2N_c-N_f}$ 
gauge theory which has $2N_c\!-\!N_f$ 
massless singlet hypermultiplets 
charged only by the $U(1)$ factors 
in addition to $N_f$ 
massless quark hypermultiplets 
belonging to the fundamental 
representations of $SU(N_f\!-\!N_c)$. 
Baryonic branch of this IR-effective theory 
exactly coincides \cite{APS} with 
that of the original microscopic 
$SU(N_c)$ theory. 
We will call the original microscopic  
and IR-effective theories  respectively 
``electric" and ``magnetic" ones.

         Let us first consider Type IIA brane 
configuration relevant to the magnetic theory. 
We need  $N_c$ D fourbranes (suspended by two 
NS fivebranes) 
to obtain the vector multiplets. 
We also need $2N_c$ D sixbranes to obtain 
the hypermultiplets. 
Notice that the $2N_c\!-\!N_f$ 
singlet hypermultiplets and 
the $N_f$ quark hypermultiplets 
in the magnetic theory 
are elementary. 
These hypermultiplets should appear in the 
light spectrum of 4-6 strings. 
To obtain the non-abelian part 
of the magnetic theory,  
$N_f\!-\!N_c$ fourbranes 
must be overlapping and 
$N_f$ sixbranes must be 
crossing these overlapping fourbranes. 
To obtain the abelian part,  
$2N_c\!-\!N_f$ fourbranes must be  
separated from each other. 
Each one of them must be crossed 
by a single sixbrane. 
The positions of the separated fourbranes 
must be invariant under the 
${\bf Z}_{2N_c-N_f}$ discrete symmetry. 
This is necessary to provide the magnetic 
description of the baryonic branch root.

     At this stage we encounter some puzzle. 
Type IIA brane configuration 
relevant to the electric description 
of the baryonic branch root consists of  
$N_c$ fourbranes and $N_f$ sixbranes. 
All the fourbranes must be overlapping and 
all the sixbranes must be crossing 
these fourbranes. 
Notice that this configuration is allowed by 
the $s$-rule.
Electric and magnetic brane configurations 
are different from each other 
although they describe the same root. 
The difference originates in the existence of 
the extra $2N_c\!-\!N_f$ sixbranes 
of the magnetic configuration. 
In the electric description 
the corresponding singlet hypermultiplets 
are considered as monopoles. 
So, they are not elementary.

\begin{figure}[t]
\epsfysize=3cm \centerline{\epsfbox{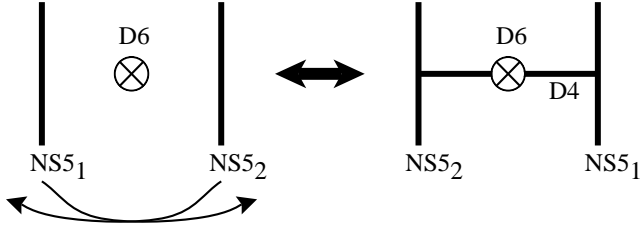}} 
\caption{\small 
 Creation and annihilation of D fourbrane by 
 D sixbrane in Type IIA picture.}
\label{BraneCreation} 
\end{figure}

\begin{figure}[t]
\epsfysize=5cm \centerline{\epsfbox{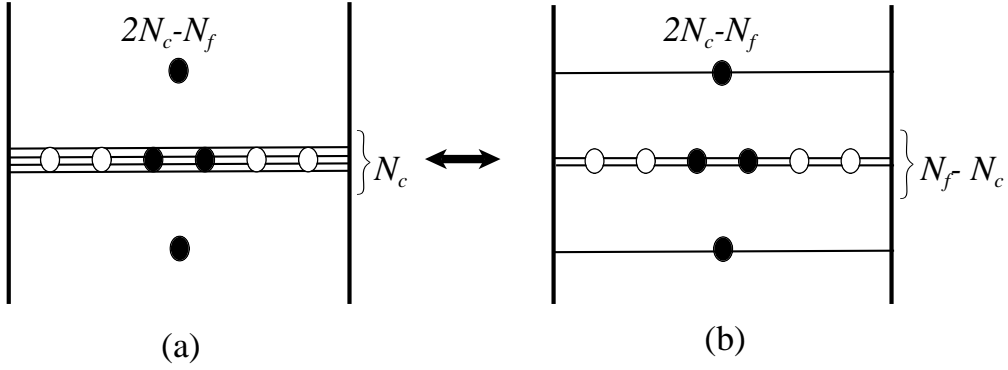}} 
\caption{\small Brane configurations related 
 by the exchange of the NS fivebranes. 
 (a) is ``electric" and (b) is ``magnetic". 
 The case of $(N_c,N_f)=(4,6)$ 
 is sketched as an example.}
\label{duality}
\end{figure}

            Let us solve this puzzle 
in Type IIA theory.  
It is argued in \cite{HW}
that, when two NS fivebranes are 
exchanged crossing a D sixbrane, 
a D fourbrane is created and 
then suspended between these fivebranes 
with touching the sixbrane. 
Conversely, 
a D fourbrane suspended between 
these fivebranes with touching 
the sixbrane is annihilated 
by this exchange. 
See Fig.\ref{BraneCreation}. 
With this process of the exchange 
the extra $2N_c-N_f$ sixbranes 
in the magnetic configuration 
annihilate the $2N_c\!-\!N_f$ fourbranes 
touching them. 
The $N_f$ sixbranes which are crossing 
the overlapping $N_f\!-\!N_c$ fourbranes 
create $N_c$ fourbranes to 
preserve the $s$-rule. 
See Fig.\ref{duality}. 
So, after the exchange, 
the magnetic configuration 
becomes very similar to 
the electric configuration. 
Finally, 
letting the extra $2N_c\!-\!N_f$ sixbranes 
go to infinity, 
we actually obtain the electric configuration.

~

       We want to make 
the above Type IIA explanation 
on the duality between 
the electric and magnetic descriptions 
of the baryonic branch root 
more rigorous by taking 
the $M$-theory viewpoint. 
In the next subsection 
we will explain the brane creation 
by considering a $M$-theory configuration 
consisting of a fivebrane and a sixbrane. 
This explanation is due to \cite{NOYY2}.
Roughly speaking, 
different boundaries of the moduli space of 
the $M$-theory configuration will give rise to 
different IIA configurations. They will be  
distinguished by the appearance of a D fourbrane. 
 In the last subsection, 
following \cite{NOY}, we present  
a family of $M$-theory configurations relevant to 
describe the brane exchange 
associated with the above duality.

\subsection{Brane creation via $M$-theory}
       
       Consider a $M$-theory configuration 
consisting of a single fivebrane 
and a single sixbrane. 
Two-dimensional part of the fivebrane 
is holomorphically embedded to the Taub-NUT space 
which the sixbrane provides. 
Let $(v,b)=(0,-b_0)$ be 
the position of the sixbrane. 
Holomorphic embedding of the fivebrane 
can be described by 
\begin{eqnarray} 
y(v,b,\sigma)= v ,
\label{single M5 in TN}
\end{eqnarray} 
where $y(v,b,\sigma)$ is 
the holomorphic coordinate of the Taub-NUT space  
\begin{eqnarray} 
y(v,b,\sigma)=
e^{-\frac{b+i\sigma}{2}}
\left( -b-b_0+\sqrt{(b+b_0)^2+|v|^2} \right). 
\end{eqnarray}

     We examine the above embedding equation 
by separating it into its angular and radial parts. 
The angular part becomes 
\begin{eqnarray} 
e^{-\frac{\sigma}{2}}=\frac{v}{|v|},  
\label{angular part}
\end{eqnarray} 
which describes a vortex of $\sigma$ 
at $v\!=\!0$. 
Recall that, 
in order to introduce the complex structure,  
we fix the Dirac string of the sixbrane 
such that it runs from $(0,-b_0)$, 
parallel with the $b$-axis, to $+\infty$. 
The vortex of $\sigma$ describes 
the intersection 
of the fivebrane and the Dirac string. 
It means that 
{\it the fivebrane is sitting to the right 
of the sixbrane in the ${\bf R}^3$ 
(or the Taub-NUT space).} 
 Fivebrane sitting to the left of the sixbrane 
can be described by 
\begin{eqnarray} 
z(v,b,\sigma)=v, 
\label{single M5 in TN2}
\end{eqnarray} 
where $z(v,b,\sigma)\equiv v/y(v,b,\sigma)$ 
\footnote{An interesting observation 
on the asymmetry between 
eqs.(\ref{single M5 in TN}) and 
(\ref{single M5 in TN2}) was made in 
\cite{BG}.}.  

\begin{figure}[t]
\epsfysize=6cm  \centerline{\epsfbox{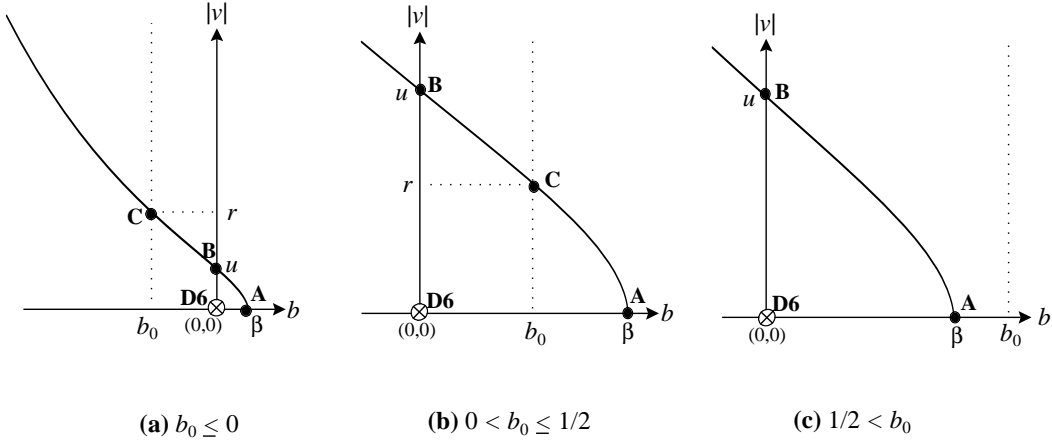}}
\caption{\small 
 Shapes of the fivebrane in the $(|v|,b)$-plane.}
\label{three cases}
\end{figure}

          The radial part of 
eq.(\ref{single M5 in TN}), 
after a parallel shift of the $b$-coordinate, 
acquires the form 
\begin{eqnarray} 
e^{-b}=e^{-b_0}
\left( b+ \sqrt{b^2+|v|^2} \right).
\label{radial part} 
\end{eqnarray}
Regarding it as an equation for $b$, 
the solution does depend only 
on $|v|$ and $b_0$.
Denote the solution by $b(|v|;b_0)$. 
Its independence from the phase of $v$ means 
that the fivebrane 
has a rotational symmetry around the $b$-axis 
in the ${\bf R}^3$. 
Differentiating eq.(\ref{radial part}) 
by $|v|$ 
one can easily find that $b(|v|;b_0)$ is 
monotonically decreasing w.r.t $|v|$. 
The shapes of the curve 
$b\!=\!b(|v|;b_0)$, 
somewhat dependent on the value of $b_0$,  
are depicted in 
Fig.\ref{three cases} 

       There exists three characteristic points 
on this curve. The first one is 
the intersection with the $b$-axis. 
The second is the intersection 
with the line $b\!=\!0$. 
The last is the intersection 
with the line $b\!=\!b_0$, 
which exists only for the case 
$b_0 \!\leq \!1/2$. 
These three points on the curve are denoted by 
A, B and C. (See Fig.\ref{three cases}.)
Their coordinates in the $(|v|,b)$-plane 
can be read from eq.(\ref{radial part}) 
\begin{eqnarray} 
\mbox{A}~:~(|v|,b)=(0,\beta(b_0)),~~~~~~~
\mbox{B}~:~(e^{b_0},0),~~~~~~~
\mbox{C}~:~(\sqrt{1-2b_0},b_0),  
\label{ABC}
\end{eqnarray} 
where we introduce the function $\beta(b_0)$ 
by $2\beta e^{\beta}\!=\!e^{b_0}$.

           We want to consider 
the IIA limit of this 
$M$-theory brane configuration. 
It will be achieved 
\cite{BIKSY} 
by taking $R\! \rightarrow \!0$ 
with fixing the value of $Rb_0/2$. 
Let us denote this fixed value by $x^6_0$. 
Recall that, 
back to the original ten-dimensional 
coordinates, the value of $Rb_0/2$ provides 
the $x^6$-position of the sixbrane. 
(The $x^{4,5}$-positions have been 
set to zero from the beginning.) 
So, this limit simply means 
making the radius of the eleven-dimension 
circle very small 
fixing the position of sixbrane.  
In this course of procedure, 
the fivebrane is always holomorphically 
embedded into the Taub-NUT space 
preserving its BPS saturation. 
So, in the limit 
we can expect to obtain 
a BPS saturated object of Type IIA theory.

\begin{figure}[t] 
\epsfysize=5cm \centerline{\epsfbox{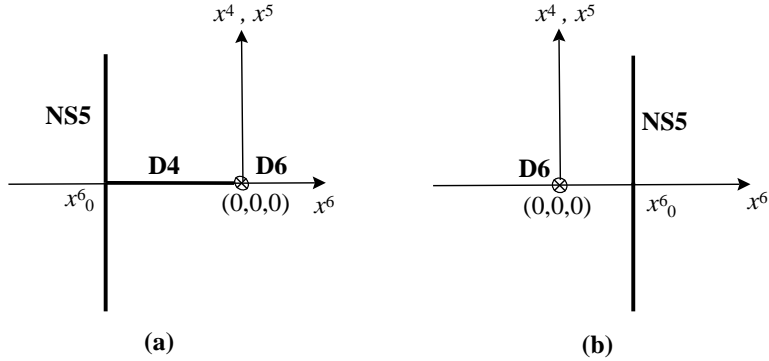}}
\caption{\small IIA limits 
 of $M$-theory brane configuration 
 depicted in Fig.\ref{three cases}: 
 (a) D fourbrane is 
 suspended between NS fivebrane and D sixbrane. 
 (b) There are only NS fivebrane and D sixbrane. }
\label{IIAlimit}
\end{figure}

                 In order to fix 
$x_0^6\!=\!Rb_0/2$ at a positive value 
for small $R$,  
$b_0$ must be a large positive quantity. 
While, 
in order to fix $x_0^6$ at a negative value 
for small $R$, 
$b_0$ must be a large negative quantity.  
Let us first examine the case of $x_0^6 \!<\!0$.  
Consider the fivebrane in the $(|v|,b)$-plane. 
It is given by the curve $b\!=\!b(|v|;b_0)$. 
Back to the original ten-dimensional coordinates, 
its coordinates become 
$\left(|x^4\!+\!ix^5|,x^6\right)=$
$\frac{R}{2}\left(|v|,b \right)$. 
We want to know how the curve 
in the original coordinates 
behaves under the limit.  
Let us pay attention to 
the characteristic points 
A,B and C on the curve. 
Quantities appearing in (\ref{ABC}) 
turn out to become as follow : 
$\mbox{lim}'_{R \rightarrow 0}R\beta(b_0)$ 
$\!=\!\mbox{lim}'_{R \rightarrow 0} Re^{b_0}$
$\!=\!\mbox{lim}'_{R \rightarrow 0} 
R\sqrt{1\!-\!2b_0}\!=\!0,$
where 
``$\mbox{lim}'_{R \rightarrow 0}$'' 
means taking 
$R \! \rightarrow \!0$ 
with fixing $x_0^6\!=\!Rb_0/2$. 
So, A,B and C become the following points on the 
$\left(|x^4\!+\!ix^5|,x^6 \right)$-plane 
\begin{eqnarray} 
\mbox{A}~:~(|x^4\!+\!ix^5|,x^6)=(0,0),~~~~~~~~
\mbox{B}~:~(0,0),~~~~~~~~
\mbox{C}~:~(0,x_0^6).   
\label{ABC in IIA 1}
\end{eqnarray} 
While this, 
one can also see that 
$\partial b(|v|;b_0)/\partial |v|$ 
goes to zero under this limit. 
Gathering all these information, 
we can conclude that 
the segment of the curve between A and C 
(which includes B) approaches to 
the straight line 
between $(0,x_0^6)$ and $(0,0)$ 
and that 
the residual part of the curve goes to 
the line $x^6\!=\!x_0^6$. 
See Fig.\ref{IIAlimit} (a). 
Owing to the rotational symmetry of the curve 
around the $x^6$-axis  
the line $x^6\!=\!x_0^6$ 
realizes a five-dimensional 
object in ten-dimensions, 
that is, 
a NS fivebrane while the line segment 
between $(0,x_0^6)$ and $(0,0)$ 
realizes a four-dimensional 
object in ten-dimensions, 
that is, a fourbrane. 
To summary, 
the IIA limit of 
the $M$-theory configuration 
with $b_0$ negative describes 
the configuration of a NS fivebrane, 
a D fourbrane and a D sixbrane. 
The NS fivebrane is located at 
$(x^6,x^7,x^8,x^9)=(x^6_0,0,0,0)$ 
while the D fourbrane with 
worldvolume $(x^0,x^1,x^2,x^3,x^6)$ 
is suspended, in the $(x^4,x^5,x^6)$-space, 
between the point $(0,0,x^6_0)$ on the NS fivebrane 
and the D sixbrane at $(0,0,0)$.

     Next let us consider the case of 
$x_0^6\!>\!0$. 
We can repeat the same argument. 
Consider the characteristic points 
A and B on the curve. 
Notice that, since we can assume $b_0!>\!1/2$ 
without losing any generality, 
the point $C$ is irrelevant for the discussion. 
Quantities appearing in (\ref{ABC}) have the 
following limits in this case : 
$\mbox{lim}'_{R \rightarrow 0}
R\beta(b_0)\!=\!2x_0^6$ 
and  
$\mbox{lim}'_{R \rightarrow 0} 
Re^{b_0}\!=\!+\infty$.  
So, A and B become the following points on the 
$\left(|x^4\!+\!ix^5|,x^6 \right)$-plane 
\begin{eqnarray} 
\mbox{A}~:~(|x^4\!+\!ix^5|,x^6)=(0,x_0^6),
~~~~~~~~~~~
\mbox{B}~:~(+\infty,0).  
\label{ABC in IIA 2}
\end{eqnarray}
Since it holds also 
$\partial b(|v|;b_0)/\partial |v| 
\! \rightarrow \!0$, 
we can conclude that the curve approaches to 
the straight line $x^6\!=\!x_0^6$. 
So, we obtain only a fivebrane.  
See Fig.\ref{IIAlimit} (b). 
To summarize, 
the IIA limit of the $M$-theory configuration 
with $b_0$ positive describes the configuration 
of a NS fivebrane and a D sixbrane. 
The NS fivebrane is located at 
$(x^6,x^7,x^8,x^9)=(x_0^6,0,0,0)$ and 
the D sixbrane is at $(x^4,x^5,x^6)=(0,0,0)$. 
There appears no D fourbrane.

~

        For the $M$-theory configuration 
(\ref{single M5 in TN}) 
with the moduli parameter $b_0$ 
we have shown that there are two possibilities of 
the IIA configuration , 
one of which is the configuration of 
a NS fivebrane, a D fourbrane and a D sixbrane 
while the other is the configuration of 
a NS fivebrane and a D sixbrane.  
Appearance or disappearance of the D fourbrane 
in the IIA configuration can be understood 
as a result of the crossing of NS fivebrane 
and D sixbrane if one takes Type IIA picture.

\subsection{Duality of baryonic branch via M-theory}

    Let us consider the following family of 
${\bf Z}_{2N_c-N_f}$-invariant 
curves $\Sigma_{\rho}$  
\begin{eqnarray} 
y^2
-2v^{N_f-N_c}
(v^{2N_c-N_f}-\Lambda^{2N_c-N_f})y 
+f(\rho)v^{N_f}
(v^{2N_c-N_f}-g(\rho)\Lambda^{2N_c-N_f})
=0 .
\label{invariant curves} 
\end{eqnarray} 
Here $\rho \in \mbox{[}0,1\mbox{]}$ 
parametrizes the family.  
$f$ and $g$ are some 
unknown functions of $\rho$. 
To determine them 
we require the following constraints 
on the curves : 
\begin{itemize}
\item As $\rho \rightarrow 0$, 
$\Sigma_{\rho}$ becomes 
the Seiberg-Witten curve 
for the electric description of 
the baryonic branch root while, 
as $\rho \rightarrow 1$,  
$\Sigma_{\rho}$ becomes 
the Seiberg-Witten curve 
for the magnetic description of 
the baryonic branch root. 
\item $\Sigma_{\rho}$ is 
completely degenerate for any $\rho$. 
\end{itemize}
These requirements are based on a naive 
M-theory generalization of the exchange of 
the NS fivebranes in Type IIA theory. 
The first one is, so to say, 
the initial and final conditions 
of the brane exchange.  
And the second one reflects 
the fact that the NS fivebranes must 
be disconnected under their exchange.

      These constraints put some restrictions 
on $f$ and $g$. 
The complete degeneration of $\Sigma_{\rho}$ 
requires  
$f$  be the following function of $g$ 
\begin{eqnarray} 
f=\frac{4(g-1)}{g^2}. 
\end{eqnarray} 
The initial and final conditions restrict 
the boundary behaviors of $g$ as  
\begin{eqnarray} 
\mbox{lim}_{\rho \rightarrow 0} g(\rho) 
=\infty ,~~~~
\mbox{lim}_{\rho \rightarrow 1} g(\rho) 
=1. 
\end{eqnarray} 
In principle we can take any function $g$ 
with these boundary behaviors. 
But, this arbitrariness is irrelevant. 
It simply means that one can move 
the fivebranes in an arbitrary manner 
as far as they exchange each other. 
We will fix it as 
\begin{eqnarray} 
g(\rho)=\frac{1}{\rho}. 
\end{eqnarray} 
With this choice of $g$ 
the factorized form of 
the curve $\Sigma_{\rho}$ becomes 
\begin{eqnarray} 
y_{+}(v)= 2(1-\rho)v^{N_c},~~~~~~~  
y_{-}(v)=2\rho v^{N_f-N_c}
(v^{2N_c-N_f}-\Lambda^{2N_c-N_f}/\rho) .
\label{factorized forms at rho}
\end{eqnarray}

         These curves are holomorphically 
embedded into the multi Taub-NUT spaces. 
The multi Taub-NUT space $Q_{\rho}$ 
in which $\Sigma_{\rho}$ is embedded 
is characterized by the positions 
of $2N_c$ sixbranes,
$(e_i,b_i)$.   
Their $v$-positions are specified 
from the equation of the curve 
\begin{eqnarray} 
e_1 = \cdots = e_{N_f}=0,~~~~ 
e_{N_f+r}(\rho)= 
\omega^r\Lambda / \rho^{\frac{1}{2N_c-N_f}}
~~~~~ 
(1 \leq r \leq 2N_c-N_f). 
\label{v-position of Qrho}
\end{eqnarray} 
We will fix their $b$-positions to 
$b_1 \!=\! \cdots \!=\!b_{2N_c}\!=\!0$.

        The holomorphic embedding 
of $\Sigma_{\rho}$ into 
$Q_{\rho}$ is given by 
\begin{eqnarray}
y_{\pm}(v)=y(v,b,\sigma). 
\label{embedding at rho} 
\end{eqnarray}  
Since the curve $\Sigma_{\rho}$ 
is completely factorized, 
each equation describes independently 
the embedding of a complex plane 
$\Sigma_{\rho +}$ or $\Sigma_{\rho -}$. 
Consider the asymptotic $b$-positions of 
$\Sigma_{\rho \pm}$, 
that is, 
the values of $b_{\pm}(v)$ 
at $v \! \approx \! \infty$. 
A simple estimation 
of eqs.(\ref{embedding at rho}) gives  
$b_{+}(v)\! \approx \! 
-\!2 \ln 2(1\!-\! \rho) \!-\!2N_c \ln |v|$ 
and 
$b_{-}(v) \! \approx \! 
-\!2 \ln 2\rho \!-\!2N_c \ln |v|$.  
Therefore their relative $b$-position 
at $v \!=\! \infty$ is 
$b_{+}(\infty)\!-\!b_{-}(\infty)
\!=\! 2 \ln \frac{\rho}{1\!-\! \rho}$, 
which is negative when 
$0 \!<\! \rho \!<\! 1/2$ 
and positive 
when $1/2 \!<\! \rho \!<\!1$. 
This means that $\Sigma_{\rho \pm}$ 
overlap asymptotically 
at $\rho \!=\! 1/2$. 
We want to propose that 
the brane exchange actually occurs 
at $\rho \!=\! 1/2$.

      To justify this proposal 
we discuss massless membranes associated with 
the fivebrane at each value of $\rho$. 
Let us begin by studying the intersection 
of $\Sigma_{\rho \pm}$ in the ${\bf R}^3$. 
Let $\pi$ be the natural projection from 
$Q_{\rho}$ to ${\bf R}^3$ as before. 
The intersection 
$\pi(\Sigma_{\rho +}) \cap \pi (\Sigma_{\rho -})$ 
is located at the $b\!=\!0$ plane in ${\bf R}^3$, 
which we identify with the $v$-plane.  
The intersection is the solution of the equation 
\begin{eqnarray} 
|y_+(v)|=|y_-(v)|.  
\end{eqnarray}
Using the explict forms 
(\ref{factorized forms at rho}) 
it is solved as 
\begin{eqnarray} 
v_{r}(\theta)
=\frac{e_{N_f+r}(\rho)}
{ \left(1+e^{i \theta}
\left| \frac{1-\rho}{\rho} 
\right| \right)^{\frac{1}{2N_c-N_f}}},
~~~~~~~
(r=1,\cdots,2N_c-N_f.
~~~
0 \leq \theta \leq 2\pi.) 
\label{lr}
\end{eqnarray} 
So, the intersection 
consists of one-dimensional lines 
$l_r$ given by 
$v\!=\!v_{r}(\theta)$

       Actual forms of the lines $l_r$ 
drastically change at  $\rho \!=\! 1/2$. 
When $0 \!<\! \rho \!<\! 1/2$, 
each line $l_r$ is connected with 
$l_{r \pm 1}$ at the boundaries. 
$l_1 \cup \cdots \cup l_{2N_c-N_f}$ 
becomes a single closed loop 
encircling $v\!=\!0$.   
On the other hand, 
when $1/2 \!<\! \rho \!<\! 1$, 
each line $l_r$ itself  
becomes a closed loop 
on the $v$-plane encircling 
$v\!=\!e_{N_f+r}(\rho)$.

                What does this qualitative 
difference of the intersection mean?  
To explain this we shall consider 
a slight perturbation 
of $\Sigma_{\rho}$. 
We slightly change the $v$-positions of 
the $2N_c\!-\!N_f$ sixbranes 
from $v\!=\!e_{N_f+r}(\rho)$. 
With this change 
the curve $\Sigma_{\rho}$ becomes  
$\tilde{\Sigma}_{\rho}$. 
This perturbed curve has genus 
$2N_c\!-\!N_f$. 
Its Riemann sheets 
$\tilde{\Sigma}_{\rho \pm}$  
are patched together by 
$2N_c\!-\!N_f$ circles $\tilde{C}_r$ 
in the multi-Taub-NUT space. 
Each $\tilde{L}_r 
\! \equiv \! \pi(\tilde{C}_r)$ is 
a one-dimensional line 
in the $v$-plane connecting 
two branch points of 
$\tilde{\Sigma_{\rho}}$ through 
$v\!=\! \omega^r \Lambda$.

            We first examine the case of 
$0\!<\! \rho \!<\! 1/2$. 
The ends of $\tilde{L}_r$ 
(two branch points of $\tilde{\Sigma}_{\rho}$) 
are getting closer to 
the ends of $\tilde{L}_{r \pm 1}$ 
as the perturbation becomes weak. 
Finally they are identified with 
the ends of $\tilde{L}_{r \pm 1}$. 
It means that 
$2N_c\!-\!N_f$ one-dimensional lines 
$\tilde{L}_r$ become 
the single closed loop 
$l_1 \cup \cdots \cup l_{2N_c-N_f}$. 
In this process 
the $\beta$-cycles 
of $\tilde{\Sigma}_{\rho}$ vanish. 
Nextly we consider the case of 
$1/2 \!<\! \rho \!<\! 1$. 
The ends of $\tilde{L}_r$ 
are getting closer to each other 
as the perturbation becomes weak.  
Finally its ends are identified 
with each other. 
So, we obtain the closed loop $l_r$. 
In this process the $\alpha$-cycles 
of $\tilde{\Sigma}_{\rho}$ vanish.

       The above consideration implies that 
the complete degeneration of $\Sigma_{\rho}$ 
is due to the vanishing $2N_c\!-\!N_f$ 
$\beta$-cycles when $0 \!<\! \rho \!<\! 1/2$ 
and due to the vanishing 
$2N_c\!-\!N_f$ $\alpha$-cycles 
when $1/2 \!<\! \rho \!<\!0$.
This reflects the qualitative difference of 
the intersection in the ${\bf R}^3$.

           Now we consider massless membranes. 
First we examine the case of 
$1/2 \!<\! \rho \!<\! 1$. 
Consider a membrane 
with worldvolume ${\bf R} \times D$, 
where $D$ is a disk and its boundary is 
on $\Sigma_{\rho}$ such that 
$\pi (D)$ is in the $v$-plane 
encircled by $l_r$. 
This membrane also touches 
the sixbrane at $v\!=\!e_{N_f+r}(\rho)$.  
We can deform this membrane  
keeping its boundary on $\Sigma_{\rho}$ 
and touching the sixbrane.  
Actually we can make the area arbitrary small. 
This is because the boundary circle of $D$ 
can shrink to a point of $\Sigma_{\rho}$ 
due to the vanishing of the $\alpha$-cycles.  
So the minimal area is zero. 
It is achieved not by a membrane  
but by a string. 
This string provides a masssless hypermultiplet. 
It is a singlet hypermultiplet 
in the magnetic description of 
the baryonic branch root. 
Next we examine the case of 
$0 \!<\! \rho \!<\! 1/2$. 
Consider the curve $\tilde{\Sigma}_{\rho}$ 
and imagine a membrane 
with worldvolume ${\bf R} \times D$ 
where $D$ is a disk with its boundary 
is one of the $\beta$-cycles of 
$\tilde{\Sigma}_{\rho}$. 
Since $\Sigma_{\rho}$ is obtained by 
vanishing the $\beta$-cycles of 
$\tilde{\Sigma}_{\rho}$, 
the area of this membrane becomes 
zero after the degeneration. 
In this limit $D$ can be regarded as a point. 
This provides a massless monopole 
in the electric description of the 
baryonic branch.


\end{document}